\RequirePackage{fix-cm}
\documentclass[smallextended]{svjour3}       
\smartqed  
\usepackage{graphicx}
\usepackage{amsmath}
\usepackage{amssymb}              

\usepackage{eurosym}

\usepackage{tikz}
\usetikzlibrary{positioning}
\usepackage{pgfplots}
\usepackage{pgfplotstable}
\usetikzlibrary{patterns}
\usepgfplotslibrary{patchplots}
\definecolor{darkgreen}{rgb}{0,0.7,0}
\definecolor{lightblue}{rgb}{0.8,0.8,1.0}
\usetikzlibrary{decorations.text}
\usetikzlibrary{decorations.pathmorphing}

\pgfplotscreateplotcyclelist{mycolor}{%
blue,every mark/.append style={fill=blue!80!black},mark=*\\%
red,every mark/.append style={fill=red!80!black},mark=square*\\%
darkgreen,every mark/.append style={fill=darkgreen!80!black},mark=otimes*\\%
black,mark=star\\%
blue,every mark/.append style={fill=blue!80!black},mark=diamond*\\%
red,densely dashed,every mark/.append style={solid,fill=red!80!black},mark=*\\%
darkgreen,densely dashed,every mark/.append style={
solid,fill=darkgreen!80!black},mark=square*\\%
black,densely dashed,every mark/.append style={solid,fill=gray},mark=otimes*\\%
blue,densely dashed,mark=star,every mark/.append style=solid\\%
red,densely dashed,every mark/.append style={solid,fill=red!80!black},mark=diamond*\\%
}

\pgfplotsset{cycle list name={mycolor}}

\pgfplotsset{
/pgfplots/bar cycle list/.style={/pgfplots/cycle list={%
{blue,fill=blue,mark=none},%
{red,fill=red,mark=none},%
{darkgreen,fill=darkgreen,mark=none},%
{black,fill=black,mark=none},%
}
},
}

\DeclareSymbolFont{bbold}{U}{bbold}{m}{n}
\DeclareSymbolFontAlphabet{\mathbbold}{bbold}

\newcommand{\R}{\mathbb{R}}
\newcommand{\N}{\mathbb{N}}
\newcommand{\Hm}{\mathcal{H}}

\newcommand{\kernel}{\phi}

\newcommand{\dotcup}{\ensuremath{\mathaccent\cdot\cup}}
\newcommand{\ps}{\mathcal{Y}}

\makeatletter
\newcommand{\vast}{\bBigg@{3}}
\newcommand{\Vast}{\bBigg@{4}}
\makeatother

\usepackage{algorithm,algpseudocode}

\journalname{Journal of Scientific Computing}
\begin{document}

\title{Algorithmic patterns for $\Hm$-matrices on many-core processors
}


\author{Peter Zaspel}


\institute{P. Zaspel \at
              Departement Mathematik und Informatik\\
	      Universtit\"at Basel\\
	      Spiegelgasse 1\\
	      4051 Basel, Switzerland\\
	      \email{peter.zaspel@unibas.ch}
}

\date{Received: date / Accepted: date}

\maketitle

\begin{abstract}
In this work, we consider the reformulation of hierarchical ($\Hm$) matrix algorithms for many-core processors with a model implementation on graphics processing units (GPUs). $\Hm$ matrices approximate specific dense matrices, e.g., from discretized integral equations or kernel ridge regression, leading to log-linear time complexity in dense matrix-vector products. The parallelization of $\Hm$ matrix operations on many-core processors is difficult due to the complex nature of the underlying algorithms. While previous algorithmic advances for many-core hardware focused on accelerating existing $\Hm$ matrix CPU implementations by many-core processors, we here aim at totally relying on that processor type. As main contribution, we introduce the necessary parallel algorithmic patterns allowing to map the full $\Hm$ matrix construction and the fast matrix-vector product to many-core hardware. Here, crucial ingredients are space filling curves, parallel tree traversal and batching of linear algebra operations. The resulting model GPU implementation \texttt{hmglib} is the, to the best of the authors knowledge, first entirely GPU-based Open Source $\Hm$ matrix library of this kind. We conclude this work by an in-depth performance analysis and a comparative performance study against a standard $\Hm$ matrix library, highlighting profound speedups of our many-core parallel approach.

\end{abstract}

\section{Introduction}
\label{sec:introduction}
In many fields of applications we are required to solve large dense linear systems of equations of the form
\begin{equation}A_{\kernel, \ps\times\ps} \vec{x} = \vec{b}\label{eq:denseLinearSystem}\end{equation}
with
\begin{equation}\label{eq:kernelMatrix} A_{\kernel,\ps\times \ps} = \left(
   \begin{array}{ccc}
     \kernel (\vec{y}_1,\vec{y}_1) & \cdots & \kernel (\vec{y}_1,\vec{y}_{N}) \\
     \vdots & \ddots & \vdots \\
     \kernel (\vec{y}_{N},\vec{y}_1) & \cdots &
     \kernel (\vec{y}_{N},\vec{y}_{N})
   \end{array}
\right)\,,\quad \vec{x},\vec{b}\in\R^{N}\,.
\end{equation}
where $\ps := \left\{\vec{y}_1,\ldots, \vec{y}_{N}\right\} \subset \Omega$ is a set of $N$ points in a space $\Omega\subset \R^d$ and $\kernel : \Omega\times\Omega\rightarrow \R$ is a bivariate \textit{kernel} function operating on that domain. 
In \textit{kernel-based interpolation} \cite{Wendland2004}, the linear system $\eqref{eq:denseLinearSystem}$ arises in the computation of interpolation coefficients.
In \textit{Gaussian Process Regression} (GPR) \cite{Rasmussen2005} kernel $\kernel$ is a covariance function and $A_{\kernel,\ps\times \ps}$ is replaced by $(A_{\kernel, \ps\times\ps} + \sigma^2 I)$ with $\sigma^2$ a (scalar) variance and $I$ the unity matrix. The same modified system also shows up in \textit{kernel ridge regression} \cite{Vovk2013}. Integral equations, discretized by e.g.~collocation, lead to similar linear systems. Note that, even though we here stick to the model problem \eqref{eq:denseLinearSystem} with collocation matrices of type \eqref{eq:kernelMatrix}, all our developments can be equally applied e.g.~in the context of boundary element method problems.
 
The problem size $N$ might get very large. As an example, $N$ could be the number of training samples in machine learning by kernel ridge regression. This can grow up to tens to hundreds of millions of samples or even more, depending on the application. At this point, obviously, linear solvers for \eqref{eq:denseLinearSystem} based on direct factorization get intractable due to their $O(N^3)$ complexity. This is overcome by iterative solvers with fast approximate dense matrix-vector product. 

In this work, we address the topic of parallelization of the fast approximate dense matrix-vector product based on \textit{hierarchical matrices ($\Hm$ matrices)} \cite{Boerm2003,Bebendorf2008,Hackbusch2015,Hackbusch2016}. Using $\Hm$ matrix techniques, a matrix-vector product for a fixed approximation accuracy is done in $O(N \log N)$ operations, given $\kernel$ is \textit{asymptotically smooth}, cf.~Section~\ref{sec:hMatrixAlgorithms}. Similar to panel clustering \cite{Hackbusch1989} and multipole techniques \cite{Greengard1997}, the core idea is to distinguish between subsets $\ps_i\times\ps_j\subset \ps\times\ps$, where $\ps_i$ and $\ps_j$ are ``close'' to each other or ``far away''. In $\Hm$ matrices, a \textit{tree-based} spatial decomposition of $\ps \times \ps$ is done. Nodes in that tree correspond to subsets of $\ps_i\times\ps_j\subset\ps \times \ps$ and thus to sub-blocks of $A_{\kernel,\ps\times \ps}$. Based on an \textit{admissibility condition}, these sub-blocks are either identified as close and thus directly evaluated or as far and thus approximated. Approximation is done either using expansions of the kernel function $\kernel$ or using low-rank approximations of the algebraically given matrix sub-block. In this work, low-rank approximations by adaptive cross approximation (ACA) \cite{Bebendorf2003} are considered, leading to a purely algebraic approach.  A further refinement of $\Hm$ matrix techniques leads to $\Hm^2$ matrices \cite{Hackbusch2000,Hackbusch2002,Boerm2004} that even exhibit $O(N)$ time complexity. Nevertheless, due to a higher algorithmic complexity, we for now stick to the classical $\Hm$ matrix techniques.

$\Hm$ matrix techniques speed up the solution process of \eqref{eq:denseLinearSystem} significantly. Nevertheless, large to huge problem sizes still cannot be tackled using a single processor core or just one workstation with a limited amount of memory. Therefore parallelization of the $\Hm$ matrix method is crucial. Parallelization of $\Hm$ matrix methods on \textit{standard processors} (CPUs) is an active research field. Research in this domain ranges from shared-memory to distributed-memory parallel $\Hm$ matrix implementations on CPUs. The results of this research are a set of parallel $\Hm$ matrix libraries, which include, but are not limited to $\Hm$-$\mbox{\texttt{Lib}}^{\mbox{\texttt{pro}}}$ \cite{Kriemann2017,Boerm2003,Kriemann2005,Grasedyck2008}, which is rather feature-complete with a shared-memory parallelization and limited distributed-memory support, \texttt{AHMED} (Another software library on hierarchical matrices for elliptic differential equations) \cite{Bebendorf} and \texttt{DMHM} (Distributed-Memory Hierarchical Matrices) \cite{Poulson} with a distributed-memory parallelization, \texttt{H2Lib} \cite{Boerm2017} with some support for shared-memory parallelism and work based on the related Hierarchically Semi-Separable (HSS) matrices \cite{Sheng2007} with the software \texttt{STRUMPACK} \cite{Rouet2016,Ghysels2016}, where the latter one is parallelized for shared- and distributed-memory. Another related, strongly CPU-parallel software for problems of type \eqref{eq:denseLinearSystem}, \eqref{eq:kernelMatrix} is PetRBF \cite{Yokota2010}. In contrast to the above works, we here address parallelization on \textit{many-core processors}. 

Many-core processors such as graphics processing units (GPUs) or Intel Xeon Phi reflect recent developments in chip production and high performance computing (HPC): Future parallel computers might show a dramatic growth in the number of parallel processing units with a strong (negative) impact on scalability of current shared-memory and distributed-memory parallelizations. Many-core processors are often assumed to be an optimal testbed for reformulations of classical algorithms towards a massive amount of parallelism, preparing for future parallel computers.

In this work, we will discuss fundamental research on new formulations of standard $\Hm$ matrix algorithms in order to expose as much parallelism as possible to many-core hardware. Our new algorithms are then implemented on a model hardware, namely GPUs (by NVIDIA). We claim that all of our algorithmic developments equivalently apply to GPU hardware of other vendors or to the Xeon Phi architecture.
There is a small set of related work for $\Hm$ matrices on many-core hardware. In \cite{Boerm2015}, the GPU-acceleration of the quadrature in a $\Hm^2$ matrix method for boundary element method problems is considered. Moreover, in \cite{Kriemann2013} many-core parallel LU-factorization for $\Hm$ matrices is presented and evaluated on a Xeon Phi device. However, these works have in common that many-core hardware is only used as an accelerator or for another computing task, and not as main computing device for the fast matrix-vector product. In contrast, we want to rely completely on many-core parallel hardware for the full $\Hm$ matrix construction and the $\Hm$ matrix-vector product. 

Other works in the field of many-core hardware concentrating on matrices of type \eqref{eq:kernelMatrix} or using other methods are the \texttt{ASKIT} library \cite{March2016} which uses GPU acceleration and some very specific tree-based approximation technique and fast multipole methods \cite{Yokota2013,Agullo2014} with e.g.~the multi-GPU parallel library \texttt{ExaFMM} \cite{Yokota2013}. While these approaches are very promising for these specific matrices, our main intention is to parallelize the entirely algebraic $\Hm$ matrix technique, allowing to be used in much more applications.

Fully relying on many-core hardware specifically requires us to provide many-core parallel reformulations of the underlying spatial data structure, the tree construction and traversal, bounding box computations and the construction and evaluation of both the dense matrix parts as well as the low-rank matrix approximations. We propose several algorithmic patterns for many-core processors in context of $\Hm$ matrices. Space filling curves, i.e.~Z order curves, are discussed as parallelized spatial data structure. This goes back to work on the fast construction and evaluation of bounding volume hierarchies on GPUs \cite{Lauterbach2009}. We use a parallel formulation of tree traversal using an array-based tree description (cf.~\cite{Merrill2012} for a background on GPU-based tree traversal). Batching or work aggregation, cf.~e.g.~\cite{Charara,Abdelfattah2017} allows to express parallelism even for code parts in which many similar non-equally sized subtasks are done, strongly optimizing bounding box calculations and low-rank approximations.

As a result of these developments, the author provides an Open Source reference implementation on GPU, which is called \texttt{hmglib} \cite{Zaspel2017}. To the best of the authors knowledge, this is the first entirely GPU-based $\Hm$ matrix library of this kind. For completeness, we should state that there is ongoing research on multi-GPU parallel hierarchical matrices in a library called \texttt{KSPARSE} \cite{Boukaram}, which is however not published and not available for download. Since very recently, there exists a preprint \cite{Boukaram2017} of the authors of \cite{Boukaram}, discussing the parallel, batched GPU-based implementation of matrix factorizations in context of hierarchical matrices. However, it does not become clear, whether the full algorithm (beyond the batched linear algebra) is performed on GPU. Moreover, the underlying code is not published. Therefore, we still claim that the proposed work is the first available entirely GPU-based $\Hm$ matrix method.

From a technical point of view, we will show that our many-core parallel model implementation on one GPU outperforms a classical \textit{sequentially} running CPU-based $\Hm$ matrix library by more than two orders of magnitude in the $\Hm$ matrix construction and by roughly one order of magnitude for the $\Hm$ matrix-vector product for a discussed model problem. Nevertheless, our main intention is to show the changes that are to be done to get an entirely many-core parallel implementation. This shall lead to a better understanding and preparation for future intrinsically extremely parallel computing hardware.

Section~\ref{sec:hMatrixAlgorithms} introduces hierarchical matrices and adaptive cross approximation. Thereafter, Section~\ref{sec:programmingModel} discusses a simplified programming model for many-core processors. This model allows to formulate many-core parallel programming patterns such as tree traversal or batching of similar sized sub-tasks. These patterns are introduced in Section~\ref{sec:programmingPatterns} and applied in Section~\ref{sec:manyCoreAlgorithms} to provide many-core parallel algorithms for $\Hm$ matrices. Section~\ref{sec:performanceResults} treats the reference GPU implementation covering an in-depth benchmark and empirical performance analysis. Finally, Section~\ref{sec:summary} concludes this work by a short summary.

\section{$\Hm$ matrix background}
\label{sec:hMatrixAlgorithms}
In the following, we will briefly summarize the necessary algorithmic and mathematic aspects of $\Hm$ matrices. This overview is partially based on \cite{Boerm2003}. For further reading see e.g.~\cite{Hackbusch2015}.

Let us start by identifying the points in $\ps=\{\vec{y}_1,\ldots,\vec{y}_N\}$ by their index set $I:=\{1,\ldots ,N\}$. A single entry $\kernel(\vec{y}_i,\vec{y}_j)$ of the system matrix $A_{\kernel,\ps\times \ps}$ corresponds an index tuple $(i,j)$. Later, we will build \textit{clusters} $\tau$, i.e.~specific subsets $\tau\subset I$. We can identify the product of two clusters, e.g.~$\tau\times\sigma\subset I\times I$, with a sub-matrix $\left. A_{\kernel,\ps\times \ps}\right|_{\tau\times\sigma}$ of the system matrix $A_{\kernel, \ps\times\ps}$. We will need this dual view between sets of index tuples and matrix entries to better understand the basic algorithmic idea of $\Hm$ matrices.

A kernel function $\kernel:\Omega\times\Omega\rightarrow\R$ is called \textit{asymptotically smooth} if there are constants $C_{as1}, C_{as2}\in\R^{>0}$ such that
$$|\partial_x^{\vec{\alpha}} \partial_y^{\vec{\beta}} \kernel(\vec{y},\vec{y}^\prime)| \leq C_{as1}(C_{as2}\|\vec{y}-\vec{y}^\prime\|)^{-|\vec{\alpha}|-|\vec{\beta}|} \vec{\alpha} + \vec{\beta} |\kernel(\vec{y},\vec{y}^\prime)|$$
for all $\vec{y},\vec{y}^\prime\in\Omega$ with $\vec{y}\neq\vec{y}^\prime$ and all multi-indices $\vec{\alpha},\vec{\beta}\in\N_0^d$. Fixing $\vec{y}\in\Omega$, the kernel evaluation $\kernel(\vec{y}, \vec{y}_{far})$ of an approximately smooth kernel function can be approximated with a controlled, small error, in case the point $\vec{y}_{far}$ is \textit{far} away from $\vec{y}$. In the $\Hm$ matrix approach, an \textit{admissibility} condition identifies matrix blocks $\left. A_{\kernel,\ps\times \ps}\right|_{\tau\times\sigma}$ that represent interactions of points with indices $\tau$ that are \textit{far} away from points with indices $\sigma$. Admissible matrix blocks are traditionally approximated via series expansions of kernel $\kernel$. We here consider the well-known alternative approach to approximate the matrix blocks by low-rank approximations as e.g.~in \cite{Bebendorf2003}. 

\subsection{Clustering}
The \textit{cluster tree} $\mathcal{T}_I=(V_I,\gamma,\mu)$ is a hierarchical spatial data structure on $I$ (or $\ps$). $V_I$ is the set of nodes in the tree, $\gamma$ a mapping $\gamma:V_I\rightarrow \mathcal{P}(V_I)$ of the nodes to their children and $\mu:V_I\rightarrow \mathcal{P}(I)$ a mapping of the nodes to their value. Here, the \textit{value} of each node is a cluster in $I$, i.e.~a subset of $I$. A cluster tree has to fulfill
\begin{description}
\item[(C1)] $\mu(v)\in \mathcal{P}(I)\setminus\{\emptyset\}$, for all $v\in V_I$,
\item[(C2)] $\mu(\mbox{root}(\mathcal{T})) = I$,
\item[(C3)] if $v\in V_I$ is a leaf, i.e.~$\gamma(v)=\emptyset$, then $|\mu(v)|\leq C_{leaf}$ and
\item[(C4)] if $v\in V_I$ is no leaf, then it has exactly two sons $\gamma(v)=\{v_1,v_2\}$ and $\mu(v)=\mu({v}_1)\,\dotcup\,\mu({v}_2)$.
\end{description}
Thereby, the cluster tree divides the full set $I$ (C2) into a hierarchy of clusters, where non-empty clusters of $I$ (represented by nodes in $\mathcal{T}_I$, C1) are disjointly partitioned into two smaller clusters (C4). In case a cluster is no longer partitioned (thus represented by a leaf), its size is bound from above by $C_{leaf}$ (C3).

In cardinality-based clustering (CBC) \cite{Boerm2003}, an algorithm to create the cluster tree decomposes the sets $\tau = \gamma(v)$ such that the subsets in the child nodes of $v$ have similar size. Moreover, the subsets shall build geometrically distinct clusters. A CBC based on space filling curves will be introduced in Section~\ref{sec:spatialDataStructure}. The splitting in the cluster tree construction is continued as long as $|\tau|>C_{leaf}$.
 
\subsection{Bounding box admissibility}
In this work, we will restrict ourselves to an admissibility condition based on bounding boxes for clusters. Other choices are possible \cite{Hackbusch2015}. For a cluster $\tau\subset I$, the bounding box $Q_\tau$ is given as
$$Q_\tau:=\prod_{i=1}^d \left[a_\tau^{(i)}, b_\tau^{(i)}\right]$$
with $a_\tau^{(i)} := \min_{j\in\tau} y_j^{(i)}$, $b_\tau^{(i)} := \max_{j\in\tau} y_j^{(i)}$ and $\vec{y}_j := \left({y}_j^{(1)},\ldots, {y}_j^{(d)}\right)^\top$. 
One admissibility condition for an index block $\tau\times\sigma\subset I\times I$ is
\begin{equation}\label{eq:admissibilityCondition}
\min\left\{\mbox{diam}(Q_\tau), \mbox{diam}(Q_\sigma)\right\} \leq \eta \mbox{dist}(Q_\tau, Q_\sigma)
\end{equation}
with $\eta\in\R^{\geq 0}$ a parameter balancing convergence and algorithmic complexity. Diameter $\mbox{diam}(Q_\tau)$ and distance $\mbox{dist}(Q_\tau, Q_\sigma)$ of bounding boxes are defined by 
$$\mbox{diam}(Q_\tau) := \left(\sum_{i=1}^d (b_\tau^{(i)} - a_\tau^{(i)})^2\right)^{1/2}\,,$$
$$\mbox{dist}(Q_\tau, Q_\sigma) := \left( \sum_{i=1}^d \left(\max\left\{0, a_\tau^{(i)}-b_\sigma^{(i)}\right\}^2 + \max\left\{0, a_\sigma^{(i)}-b_\tau^{(i)}\right\}^2\right)\right)^{1/2}\,.$$
\subsection{Block cluster tree}

\begin{algorithm}[t]
  \caption{Algorithm to build a block cluster tree}\label{alg:buildBlockTree}
  \begin{algorithmic}
    \Procedure{build\_block\_cluster\_tree}{$v_1$, $v_2$, $w$, $C_{leaf}$}
	\State $(\tau, \sigma) \gets (\mu(v_1), \mu(v_2))$
	\If{$\tau\times\sigma$ is not admissible and $|\tau|>C_{leaf}$ and $|\sigma|>C_{leaf}$}
	\State $\gamma(w)\gets \emptyset$
	\For{$v_1^\prime\in\gamma(v_1)$} \Comment{Loop over all combinations of children in both cluster trees.}
		\For{$v_2^\prime\in\gamma(v_2)$}
			\State $\mu(w^\prime) \gets \mu(v_1^\prime)\times \mu(v_2^\prime)$ \Comment{Set block cluster of new node $w^\prime$.}
			\State $\gamma(w) \gets \gamma(w) \cup \{w^\prime\}$ \Comment{Add new node to children of $w$.}
			\State \Call{build\_block\_cluster\_tree}{$v_1^\prime$, $v_2^\prime$, $w^\prime$, $C_{leaf}$}
		\EndFor 
	\EndFor
	\Else
		\State $\gamma(w) \gets \emptyset$ \Comment{No child nodes are created, i.e.~$w$ becomes a leaf.}
	\EndIf 
    \EndProcedure
  \end{algorithmic}
\end{algorithm}

A hierarchy over \textit{blocks} $\tau\times\sigma\subset I\times I$ is induced by the \textit{block cluster tree} $\mathcal{T}_{I\times I}=(V_{I\times I}, \gamma, \mu)$, with $\gamma$ the child node map and $\mu: V_{I\times I} \rightarrow \mathcal{P}(I\times I)$ the map of nodes to their values, i.e.~blocks. Note that we re-use here the same notation ($\gamma$, $\mu$) as for the cluster tree. Algorithm~\ref{alg:buildBlockTree} implicitly defines the block cluster tree. For given cluster tree nodes $v_1,v_2$ (corresponding to clusters $\mu(v_1)=\tau, \mu(v_2)=\sigma$), a block cluster tree node $w$ with $\mu(w):=\mu(v_1)\times \mu(v_2)$ (corresponding to $\mu(w)=\tau\times\sigma$) and parameter $C_{leaf}$, this algorithm recursively constructs a block cluster tree. Procedure \texttt{build\_block\_cluster\_tree} is initially launched with $v_1$ and $v_2$ each being a root of the cluster tree $\mathcal{T}_I$ and node $w$ is initialized to represent the index block $I\times I$. By construction, the leafs of $\mathcal{T}_{I\times I}$, namely $\mathcal{L}_{I\times I} := \{ w\in V_{I\times I}\,|\,\gamma(w)=\emptyset\}$, correspond to index blocks that form a partition of $I\times I$.

\subsection{\textbf{R}$k$-matrices and adaptive cross approximation}
If a node $w$ in a block cluster tree corresponds to an index block $\tau\times\sigma\subset I\times I$ that is admissible, the corresponding sub-matrix $\left. A_{\kernel,\ps\times \ps}\right|_{\tau\times\sigma}\in\R^{|\tau| \times |\sigma|}$ is replaced by an \textit{$\mathbf{R}k$ matrix} $R_{\tau\times\sigma}\in\R^{|\tau| \times |\sigma|}$. An $\mathbf{R}k$ matrix $R_{\tau\times\sigma}$ is given as
$$R_{\tau\times\sigma}=U_{\tau\times\sigma} V_{\tau\times\sigma}^\top,\quad U_{\tau\times\sigma}\in\R^{|\tau|\times k}, V_{\tau\times\sigma}\in\R^{|\sigma|\times k}\,,$$
that is, it has a maximum rank of $k$. Moreover, using $U_{\tau\times\sigma}$ and $V_{\tau\times\sigma}$, a matrix-vector product involving $R_{\tau\times\sigma}$ can be computed in $O\left(r\cdot (|\tau|+|\sigma|)\right)$ operations.

While there are many (problem-dependent) ways to approximate\linebreak $\left. A_{\kernel,\ps\times \ps}\right|_{\tau\times\sigma}$, we here aim at using a purely algebraic low-rank approximation method to derive $R_{\tau\times\sigma}$. Our method of choice is the \textit{adaptive cross approximation} (ACA) \cite{Bebendorf2003,Bebendorf2009}. This method builds a low-rank approximation by an iterative rank-one update process that is terminated based on  the eror $\epsilon$ in the Frobenius norm $\|\cdot \|_F$. 

\begin{algorithm}[t]
  \caption{Adaptive cross approximation (ACA) for $A\in\R^{m\times n}$ \cite{Bebendorf2009}\cite{Bebendorf2003}}\label{alg:aca}
  \begin{algorithmic}
    \Function{compute\_adaptive\_cross\_approximation}{$A$, $\epsilon$}
	\State $k_{max} \gets k$
	\For{$r=1,2,\ldots, k$}
		\State $\hat{\vec{u}}_r = A_{1:m,j_r} - \sum_{l=1}^{r-1} \vec{u}_l(\vec{v}_l)_{j_r},$ \Comment{Col.~index $j_r$ depending on implementation}
		\State $\vec{u}_r = (\hat{\vec{u}}_{i_r})^{-1} \hat{\vec{u}}_r,$ with $|(\hat{\vec{u}}_r)_{i_r}| = \|\hat{\vec{u}}_r\|_\infty$ \Comment{Row index $i_r$ given as pivot position} 
		\State $\vec{v}_r = \left(A_{i_r, 1:n}\right)^\top - \sum_{l=1}^{r-1} (\vec{u}_l)_{i_r} \vec{v}_l$ 
		\If{$\left(\|\vec{u}_r\|_2 \|\vec{v}_r\|_2 \leq \frac{\epsilon (1.0-\eta)}{1.0+\epsilon} \left\|\sum_{l=1}^r \vec{u}_l\vec{v}_l\right\|_F\right)$}\Comment{Stopping criterion}
			\State $k_{max} \gets r$ \Comment{$k_{max}$ is adaptively found rank}
			\State stop loop
		\EndIf
	\EndFor
	\State $U \gets (\vec{u}_1,\ldots, \vec{u}_{k_{max}})$
	\State $V \gets (\vec{v}_1,\ldots, \vec{v}_{k_{max}})$
	\State \Return $U$, $V$
    \EndFunction
  \end{algorithmic}
\end{algorithm}

One version of adaptive cross approximation is given in Algorithm~\ref{alg:aca}. It follows the lines of \cite{Bebendorf2009}. The algorithm computes for a general matrix $A\in\R^{m\times n}$ and error threshold $\epsilon$ matrices $U\in\R^{m\times k_{max}}$, $V\in\R^{n\times k_{max}}$ such that $A\approx UV^\top$. In case the algorithm terminates due to the stopping criterion, $k_{max}$ becomes the (adaptively computed) rank such that $\|A-UV^\top\|_F \leq \epsilon$.  Otherwise, the maximum rank of $k_{max}$ is hit. The choice of a column pivot index $j_r$ is strongly problem-dependent. For simplicity, we choose $j_r$ such that $\|\hat{\vec{u}}_r\|_2>\epsilon_{0}$ for small $\epsilon$ in the range of machine precision. In our practical implementation, we will, however, avoid to evaluate the stopping criterion and will only impose the maximum rank $k_{max}$. As we will see in Section~\ref{sec:convergenceResults}, $k_{max}$ can be chosen rather small due to the exponential convergence of ACA for appropriate kernel functions $\kernel$. For more details on ACA, see \cite{Bebendorf2009,Bebendorf2003}.

\subsection{$\Hm$-matrices and their matrix-vector product}
Formally, a general matrix $L\in\R^{|I|\times |I|}$ is --- for fixed $k\in\N$ and block cluster tree $\mathcal{T}_{I\times I}$ --- called \textit{$\Hm$ matrix} of blockwise rank $k$, if 
$$\mbox{rank}(\left. L\right|_{\tau\times\sigma}) \leq k$$
for all index blocks $\tau\times\sigma$ in admissible leafs. The operation to transform an existing dense matrix, e.g.~$A_{\kernel,\ps\times \ps}$, to $\Hm$ matrix form is called \textit{truncation}. It involves the introduction of a cluster tree $\mathcal{T}_I$, a block cluster tree $\mathcal{T}_{I\times I}$ and the computation of a low-rank approximation of matrix blocks corresponding to admissible leafs.

\begin{algorithm}[t]
  \caption{Matrix-vector product with an $\Hm$ matrix $L\in^{I\times I}$}\label{alg:hMatrixMatVec}
  \begin{algorithmic}
    \Function{matrix\_vector\_product}{$L$, $w$, $x$, $z$}
	\If{$\gamma(w)\neq \emptyset$}
		\For{$w^\prime\in\gamma(w)$}
			\State \Call{matrix\_vector\_product}{$L$, $w^\prime$, $x$, $z$}
		\EndFor
	\Else
		\State $\tau\times\sigma \gets \mu(w)$
		\If{$\tau\times\sigma$ is admissible}
			\State $\vec{t}\gets V_{\tau\times\sigma}^\top \left. \vec{x}\right|_\tau$
			\State $\left. \vec{z}\right|_\tau \gets \left. \vec{z}\right|_\tau + U_{\tau\times\sigma} \vec{t}$
		\Else
			\State $\left. \vec{z}\right|_\tau \gets \left. \vec{z}\right|_\tau + \left. L\right|_{\tau\times\sigma} \left. \vec{x}\right|_\sigma$
		\EndIf
	\EndIf
	\State \Return $z$
    \EndFunction
  \end{algorithmic}
\end{algorithm}

The (fast) matrix-vector product of an $\Hm$ matrix $L\in\R^{|I|\times |I|}$ with a vector $\vec{x}\in\R^{|I|}$, that is, the efficient evaluation of
$$\vec{z}:=\vec{z} + L \vec{x}\,,$$
is summarized in Algorithm~\ref{alg:hMatrixMatVec}. The algorithm recursively traverses the block cluster tree for an initially given (root) node $w$ and applies a low-rank matrix-vector product for admissible blocks and the full dense matrix for non-ad\-mis\-si\-ble blocks. If we launch \textsc{matrix\_vector\_product} with $w$ corresponding to $I\times I$ and $L$ being the truncated version of $A_{\kernel,\ps\times \ps}$, it can be shown that the algorithm has a complexity of $O(k \cdot N \log{N})$ \cite{Hackbusch2015}.

\section{Programming model for many-core parallel algorithms}
\label{sec:programmingModel}

In this Section, we introduce the terminology to describe efficient and scalable parallel many-core algorithms. Note that, to the best of the author's knowledge, a common abstract programming model for many-core architectures is still missing. Therefore, algorithmic work on GPUs or Xeon Phi often addresses many details of these architectures. In contrast, we use a strongly simplified programming model, avoiding most of the technical details of classical many-core literature. Our model is based on two observations. First, a crucial part of a lot of many-core parallel algorithms requires almost no interaction between the involved parallel compute units, that is, they are close to \textit{embarrassingly} parallel. Second, vendors (or enthusiasts) provide extremely efficient many-core parallel implementations of base algorithms (reductions, scan operations, etc.) for more complex parallel algorithmic patterns. Therefore, we claim that we can build all algorithms of interest by combinations of almost embarrassingly parallel kernels and standardized parallel algorithms. They are defined in more detail in the following paragraphs.

\subsection{Almost embarrassingly parallel kernels}
\label{sec:kernels}
The kind of compute kernels we discuss here are strongly related to the \textit{bulk synchronous parallel model}, cf.~\cite{Valiant1990}: We introduce an (in principle) infinite number of virtual parallel threads. In each parallel thread, the same piece of sequential code is executed. Different memory accesses / execution paths are realized by a thread index which is associated to each thread. 

All threads are aggregated in a \textit{kernel}, which gets the number of threads to execute at launch time. The kernel terminates when all threads have stopped the execution of the sequential code. The sequential code (per thread) can either use \textit{local memory}, which can only be read by that single thread, or \textit{global memory}, which is available to all threads. At the end of the kernel execution, all local memory data is lost while global memory entries remain available. Whenever a single thread writes to a given global memory entry, read or write operations on that memory entry (by another thread) are invalid / prohibited. Reading (without writing) from a common global memory location by multiple threads in one kernel is possible.

One exception to the write rule is available in case of \textit{atomic} operations (usually \textit{atomic\_add} or \textit{atomic\_compare\_and\_swap}) on global memory. Atomic operations issued by different threads on one common global memory location are all correctly executed, even if this meas that threads get serialized. However, the ordering of the execution is not assured. Therefore, atomic operations are only useful in very few cases (e.g.~counters). 

Note that the actual mapping of threads to hardware processing units is not part of the model. This especially allows to define parallel programs independent of the number of available hardware threads. Moreover, the beforehand given definition of computing kernels does not give any hints towards the performance of their actual mapping to a given hardware platform. Let us give examples for GPUs. Here, global memory accesses are fast if they are done consecutively for consecutive thread indices, that is, threads 0,1,2,3,\ldots~access memory entries $e, e+1, e+2, e+3, ...$\,. In contrast, random access has rather low performance. Moreover, conditionals in the thread-sequential code of a kernel might have a severe impact on performance on GPUs if thread execution paths diverge. Other architectures might have similar limitations.

\subsection{Standardized parallel algorithms}
\label{sec:standardizedParallelAlgorithms}
As second ingredient to our many-core parallel algorithms, we expect to have access to a parallel library of standardized (many-core parallel) algorithms similar to the C++ Standard Template Library (STL) algorithms library.  We e.g.~need \texttt{reduce}, \texttt{stable\_sort}, \texttt{scan}, ... These algorithms are expected to be realized as one function call that is executed on data in global memory. The many-core parallel implementation of these algorithms is assumed to be extremely optimized and given e.g.~by the vendor. On GPUs an implementation of STL-like algorithms is available via the \texttt{Thrust} library \cite{Bell2011}. Alternatives include, but are not limited to \texttt{ArrayFire} \cite{Yalamanchili2015} (supporting GPUs, CPUs and Xeon Phi) and \texttt{Boost.Compute} \cite{Szuppe2016} (supporting multi-core CPUs and GPUs). In addition, we assume to have appropriate BLAS libraries for a given many-core device.

\section{Many-core parallel programming patterns for $\Hm$ matrices}
\label{sec:programmingPatterns}
As motivated before, we introduce in the following a set of parallel programming patterns that are necessary for algorithms based on $\Hm$ matrices.

\subsection{Parallel tree traversal}

\begin{figure}[t]
\begin{center}
\scalebox{0.2}{\includegraphics{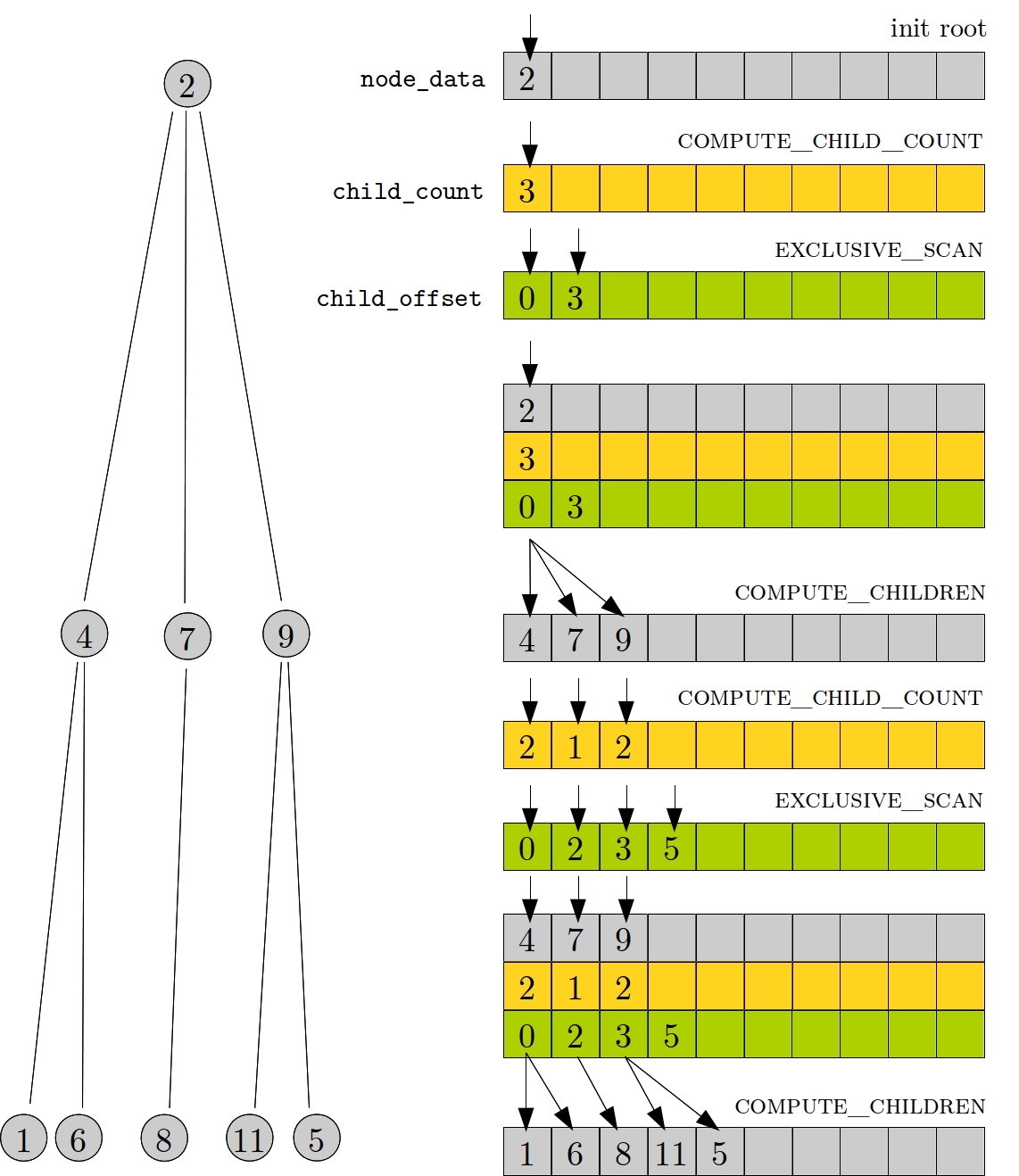}}
\end{center}
\caption{\label{fig:treeTraversal}The many-core parallel tree traversal parallelizes over the nodes on a given level of the tree. Our algorithm is here exemplified for nodes containing numbers.}
\end{figure}

\begin{algorithm}[t]
  \caption{Many-core parallel tree traversal}\label{alg:treeTraversal}
  \begin{algorithmic}
    \Procedure{traverse}{root}\Comment{Traverse a tree with given root data}
	\State allocate {node\_data}, {node\_data\_old}, {child\_count}, {child\_offset}
	\State {node\_data[0]} $\gets$ {root\_data}
	\State $l \gets 0$
	\State $|V(l)| \gets 1$
	\While{$|V(l)|>0$} \Comment{Handle tree levels as long a there are nodes}
		\State \Call{compute\_child\_count<$|V(l)|$>}{child\_count, node\_data}
		\State \Comment{Problem-dependent kernel to compute the number of children per node}
		\State \Call{exclusive\_scan}{child\_offset, child\_count, $0$, $|V(l)|$}
		\State $|V(l+1)| \gets $ child\_offset[$|V(l)|$]\Comment{Set total number of children of next level}
		\State node\_data\_old $\gets$ node\_data
		\State \Call{compute\_children<$|V(l)|$>}{node\_data, node\_data\_old, child\_count, child\_offset}
		\State \Comment{Problem-dependent kernel to compute the content of the children}
		
		\State $l \gets l+1$
	\EndWhile
    \EndProcedure
  \end{algorithmic}
\end{algorithm}

In the following, we introduce a fully parallel tree traversal algorithm, which is related to ideas in \cite{Lauterbach2009,Merrill2012}. It on-the-fly builds and traverses a tree. The tree traversal algorithm is given in Algorithm~\ref{alg:treeTraversal}. An input tree $\mathcal{T}=(V,\gamma,\mu)$ is assumed to have height $h(\mathcal{T})$, levels $l\in\{0,\ldots, h(\mathcal{T})\}$ and nodes $v\in V$ of arbitrary order. The algorithm is designed such that we only store nodes $V(l):=\{v\in V\,|\, level(v)=l\}$ and $V(l+1)$ for two consecutive levels $l$ and $l+1$. All other data is created level-wise and discarded after a new level has been successfully created. The nodes $v\in V(l)$ and $v^\prime\in V(l+1)$ are stored in global arrays \texttt{node\_data\_old} and \texttt{node\_data}. In addition, we need for level $l+1$ the number of children per node $|\gamma(v^\prime)|$ (stored in \texttt{child\_count}) and the offset of the data of the child nodes (\texttt{child\_offset}). Figure~\ref{fig:treeTraversal} illustrates these arrays.

The algorithm works as follows: Let us assume for now that the arrays per level can have arbitrary size and that we are on level $0\leq l<h(\mathcal{T})$ and the only available data is the node data. We first invoke a kernel \texttt{compute\_child\_count} with the number of threads equal to the number $|V(l)|$ of nodes on that level, thus the number of valid entries in the \texttt{node\_data} array. In each thread, we independently compute for each node $v\in V$ (based on the node data) the number of children $|\gamma(v)|$ that shall be created on the next level. This computation is problem-dependent. In case of the cluster tree, it e.g.~holds $|\gamma(v)|\in\{0,2\}$. The number of children is stored at the same offset in the array as the given node data. In a next step, we have to compute the offsets for the node data on the next level, i.e.~\texttt{child\_offset}. This can be done by an \texttt{exclusive\_scan} operation initialized to $0$. The entries of \texttt{child\_offset} then become \texttt{[0, child\_count[0], child\_count[0]+child\_count[1],\ldots]}. The output of the scan operation contains as additional number (at the end of the set of valid entries) the total number of children $|V(l+1)|$. The last step on level $l$ is the creation of the node data $V(l+1)$ on level $l+1$. This is again done using a kernel with the number of threads equal to $|V(l)|$. Each thread then independently computes the new entries taking the storage location in \texttt{node\_data} for level $l+1$ from \texttt{child\_offset}. This finishes the computation for one level. The whole process is iteratively proceeded over all levels $0\leq l<h(\mathcal{T})$. To start the tree traversal on level zero, i.e.~the root of the tree, the \texttt{node\_data} array is initialized with a single entry. A full example of a tree traversal is given in Fig.~\ref{fig:treeTraversal} and the algorithm is stated in Algorithm~\ref{alg:treeTraversal}.

We next have to discuss how to deal with the array allocation, knowing that the required size of the storage arrays differs between the tree levels.  Here, we have two options. The first option would be to pre-allocate the arrays to a fixed size $\max_{l\in\{0,...,height(\mathcal{T})\}} |V(l)|$. This, of course, requires to know this number beforehand or to have a suitable upper bound for it. Very often, this is not the case. The second option is a \textit{dynamic} allocation of the array size for the next level. This size can be predicted based on the information in the \texttt{child\_count} array. In case a reallocation of memory is a very expensive operation on a given target architecture, one could also apply hybrid approaches such as adapting the size of the arrays only if a given array (of large size) would be too small for the next level. In our implementation on GPU, a global reallocation of the memory is a very efficient operation. This is why we have chosen to use the dynamic allocation approach.

Finally, we should have a look at the properties of the algorithm in terms of the use of the many-core processor. It becomes obvious that the number of utilized parallel threads on the first few levels is very low. That is, the proposed algorithm makes no full use of the many-core processor on the first levels. This might become an issue if many tree traversals on small trees are considered and if the tree traversal operation itself is the dominant operation in an application. However, both is not the case in our application: The trees are very large and, as we will see in Section~\ref{sec:performanceResults}, the tree traversal operation makes only a very small fraction of the overall $\Hm$ matrix setup / application process. Therefore, we consider our tree traversal method efficient enough for our needs. In case higher utilization of the many-core processor is needed, efficient solutions become very architecture-specific. In case of GPUs, there is work on tree traversal by work queues \cite{Garanzha2011}, which, however, makes explicit use of knowledge on the hardware and which somehow even breaks the programming model initially considered for GPUs.

\subsection{Batching many similar non-equally sized compute tasks}
\label{sec:batching}
\begin{figure}[t]
\begin{center}
\scalebox{0.1}{\includegraphics{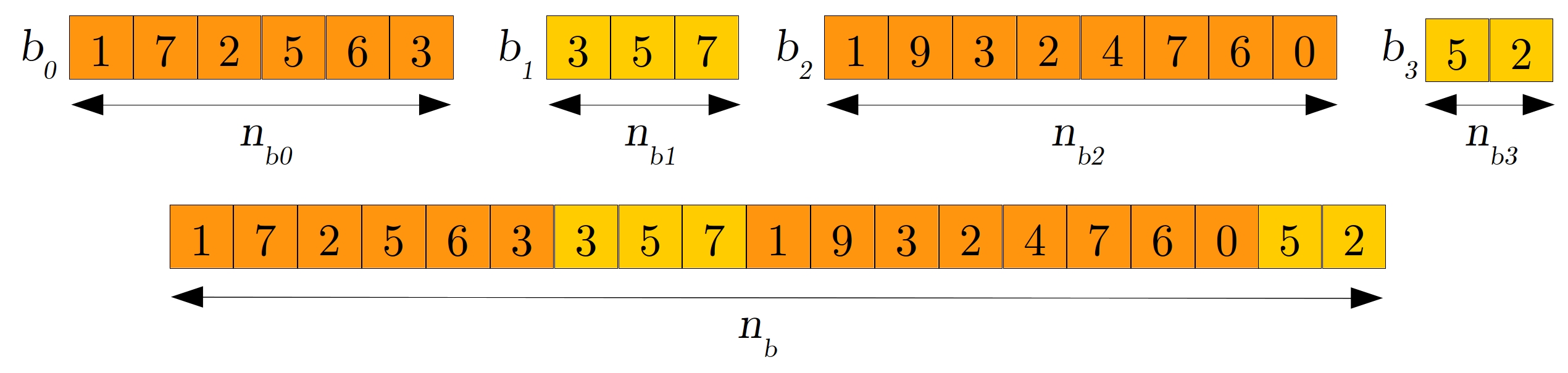}}
\end{center}
\caption{\label{fig:batchingExample}By batching individual subproblems into one big array, it becomes possible to utilize a many-core processor much better.} 
\end{figure}

We next want to discuss how to make optimal use of a many-core processor in case there is an identical computing task which shall be applied to $m$ different, non-equally sized arrays $b_0, b_1,\ldots, b_{m-1}$ of sizes $n_{b_0}, n_{b_1}, \ldots, n_{b_{m-1}}$. Figure~\ref{fig:batchingExample} gives an example of such arrays. The easiest way to consider a parallelization on many-core hardware would be to loop over all arrays $b_i$ and to perform the necessary many-core parallel operations individually to each array. This is efficient as long as the many-core processor is sufficiently utilized. However, we here consider arrays of changing and usually small size. In this case, a major part of the many-core processor is not used. Therefore we propose to use the technique of \textit{batching} of the necessary computations, cf.~\cite{Charara,Abdelfattah2017}, in order to use the full processor while speeding up the calculation.

The first step in batching is to put all sub-arrays or \textit{batches} $b_i$ consecutively in a \textit{batched array} of size $n_b := \sum_i n_{b_i}$, cf.~Fig.~\ref{fig:batchingExample}. We next have to distinguish between \textit{transformation} operations and \textit{reduction} operations on that batched array. A transformation on each batch applies changes individually to each entry of each batch, i.e.~there is no interaction between the data entries. Applying a transformation to each batch is therefore equivalent to applying the same transform to the full batched array. Therefore, in case of transformations, we apply one operation to the full batched array.
 
\begin{figure}[t]
\begin{center}
\scalebox{0.1}{\includegraphics{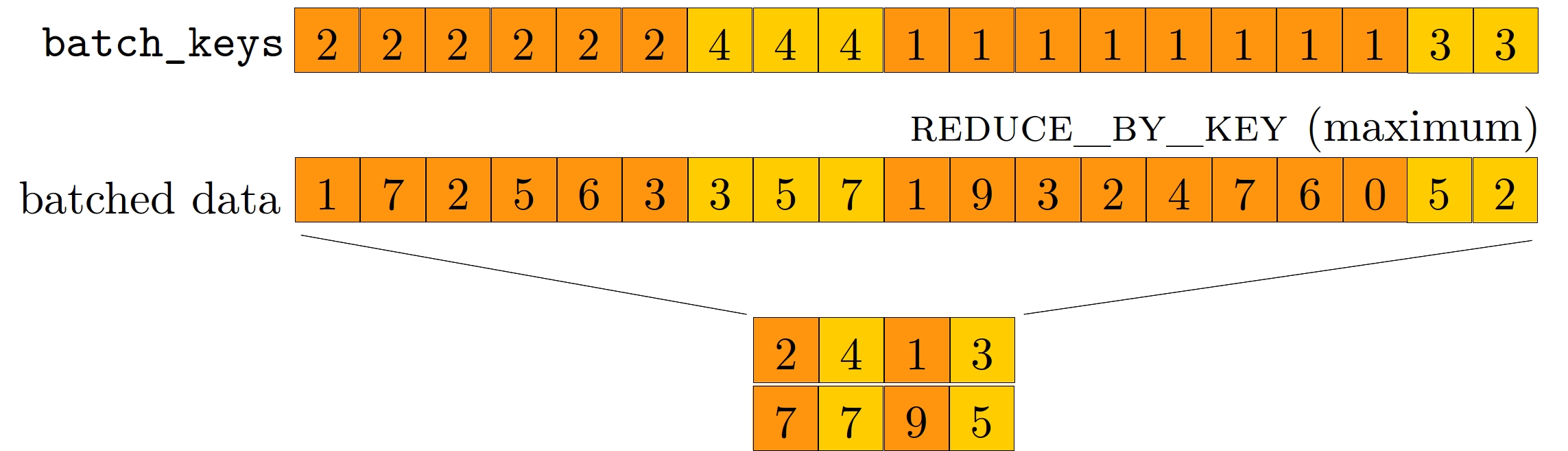}}
\end{center}
\caption{\label{fig:boundingBoxesReduceByKey}Reduction operations (e.g.~maximum computations) for several sub-problems can be handled in parallel by a \texttt{reduce\_by\_key} operation, where identical consecutive entries in \texttt{batch\_keys} mark the individual sub-problems.}
\end{figure}

In contrast, reduction operations (such as sum, minimum, maximum, norm, etc.)~require the interaction of all entries within a batch. Therefore, we need a different strategy. The STL-type algorithm \texttt{reduce\_by\_key} is applied to the full batched array and computes, in parallel, batch-wise reductions. The action of the method is shown in Fig.~\ref{fig:boundingBoxesReduceByKey} for a maximum reduction operation. We introduce a \texttt{keys} array of integer values. A series of identical numbers in the \texttt{keys} array marks one batch. The method \texttt{reduce\_by\_key} then applies the reduction operation per subset and builds up a small array of size $m$ containing the reduction results and the keys reduced to a single number. 

\begin{figure}[t]
\begin{center}
\scalebox{0.1}{\includegraphics{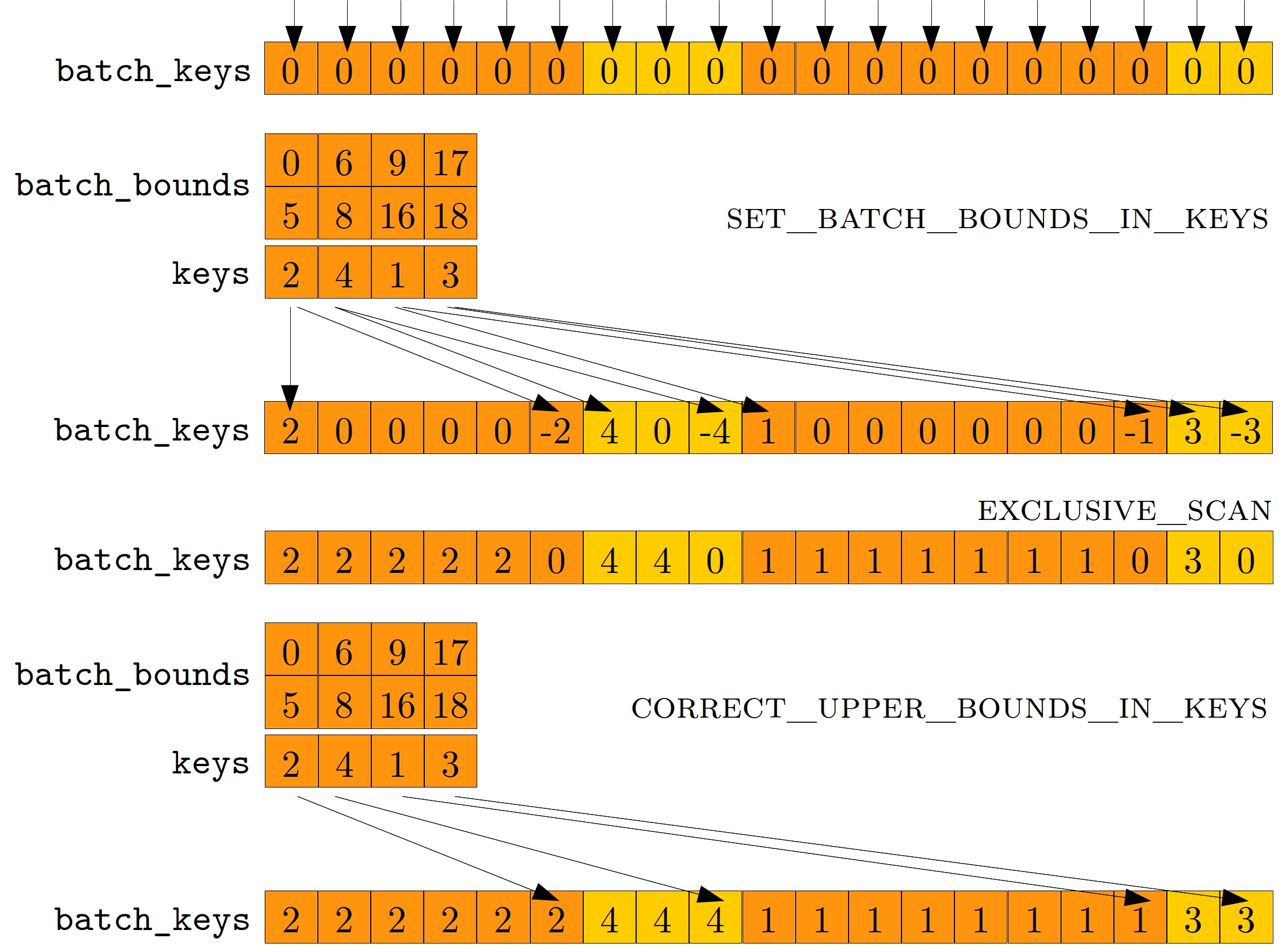}}
\end{center}
\caption{\label{fig:keysComputation}The construction of an array of keys for batching involves marking boundaries of the batches and an \textsc{exclusive\_scan} operation.}
\end{figure}

\begin{algorithm}[t]
  \caption{Many-core parallel key generation for batching}\label{alg:batchingKeys}
  \begin{algorithmic}
    \Procedure{create\_keys}{batch\_bounds, batch\_keys, $n_b$, $m$}
	\State \Call{init<$n_b$>}{keys, $0$}
	\State \Call{set\_batch\_bounds\_in\_keys<$m$>}{keys, batch\_bounds, batch\_key}
	\State \Call{exclusive\_scan}{keys, keys, $0$, $n_b$} \Comment{Write exclusive scan on keys to keys}
	\State \Call{correct\_upper\_bounds\_in\_keys<$m$>}{keys, batch\_bounds, batch\_key}
	\State \Return keys
    \EndProcedure
  \end{algorithmic}
\end{algorithm}

To compute the keys, we need an additional parallel algorithm, cf.~Algorithm~\ref{alg:batchingKeys}. It takes an array of boundaries (\texttt{batch\_bounds}) of each batch $b_i$ and an array (\texttt{batch\_keys}) of keys $k_{b_i}$ per batch as input. The procedure to create keys for batching is exemplified in Fig.~\ref{fig:keysComputation}. We initialize (by a kernel of $n_b$ threads) the \texttt{keys} array to zeros. Then, the kernel \textsc{set\_batch\_bounds\_in\_keys} of $m$ threads is invoked, where each thread independently writes the key $k_{b_i}$ and the negative key $-k_{b_i}$ to the lower and upper bound of each batch in the batched array, cf.~Fig.~\ref{fig:keysComputation}. Then an \texttt{exclusive\_scan} operation (adding elements) is executed on the full batched array. This sets the correct keys almost everywhere, except at the upper boundary of each batch. Therefore a second kernel of $m$ threads is invoked to correct the upper bounds of each batch $b_i$ to $k_{b_i}$.

In some cases, the size $n_b$ of the batched array is too large to be kept in the memory of the many-core processor. Such cases can be handled by appropriately partitioning the batches $b_i$ into subsets of batches which are then handled as before. 

A crucial property of the approach presented here is its independence of the size and the number of batches. The runtime for this approach is almost constant with the size $n_b$. This is a strong advantage over strategies that directly rely on the use of the different parallelization hierarchies (\textit{thread blocks}, \textit{grids} on GPUs and \textit{vectorization}, \textit{shared-memory parallelism}, etc.~on Xeon Phi).

\subsection{Parallel output queues}
\label{sec:parallelOutputQueues}
In some cases, we need to create what we define as \textit{write-only parallel output queues}. Such queues can only be \textit{filled} (in parallel). Removal of data or reading the head of the queue during the enqueueing process is not required. Instead, the stored queue data is handled as one array as post-processing step. As an example for such a queue, let us consider a parallel tree traversal in which (unordered) tasks are created in each leave. Instead of executing the task during the tree traversal, we can, in parallel, put them in a queue. The actual execution of the tasks can be issued afterwards as new parallel operation.

\begin{figure}
\begin{center}
\scalebox{0.1}{\includegraphics{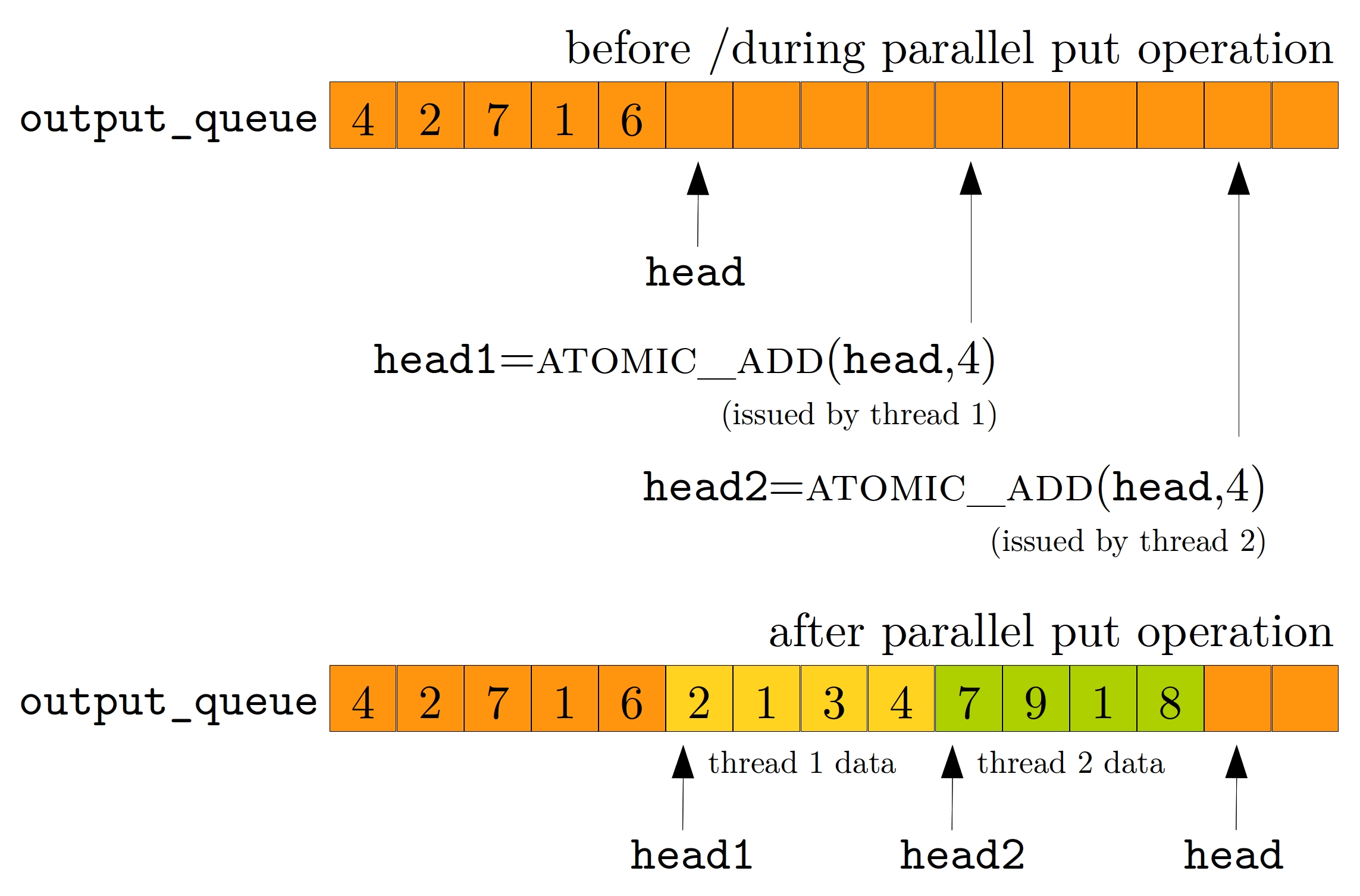}}
\end{center}
\caption{\label{fig:outputQueue}By the use of atomic operations it is easily possible to create a write-only parallel output queue. In the above example, two threads concurrently add four entries to the head of the queue.}
\end{figure}

The implementation of our parallel output queue, relies on an underlying global memory output array of appropriate size. If we cannot predict the size, we can also apply dynamic memory allocation approaches, as above. We store a pointer to the head and the tail of the queue in global memory. Whenever a put operation is issued in a thread of a kernel, the head pointer is moved accordingly by an atomic operation while storing the old head in the same operation. The old head is used as output address to write the data in the queue. Figure~\ref{fig:outputQueue} summarizes and exemplifies the approach.

\subsection{Spatial data structure by Z-order curves}
\label{sec:spatialDataStructure}

\begin{figure}
\,\vspace*{-6em}
\begin{center}
\hspace*{-3em}\scalebox{0.28}[0.3]{\includegraphics{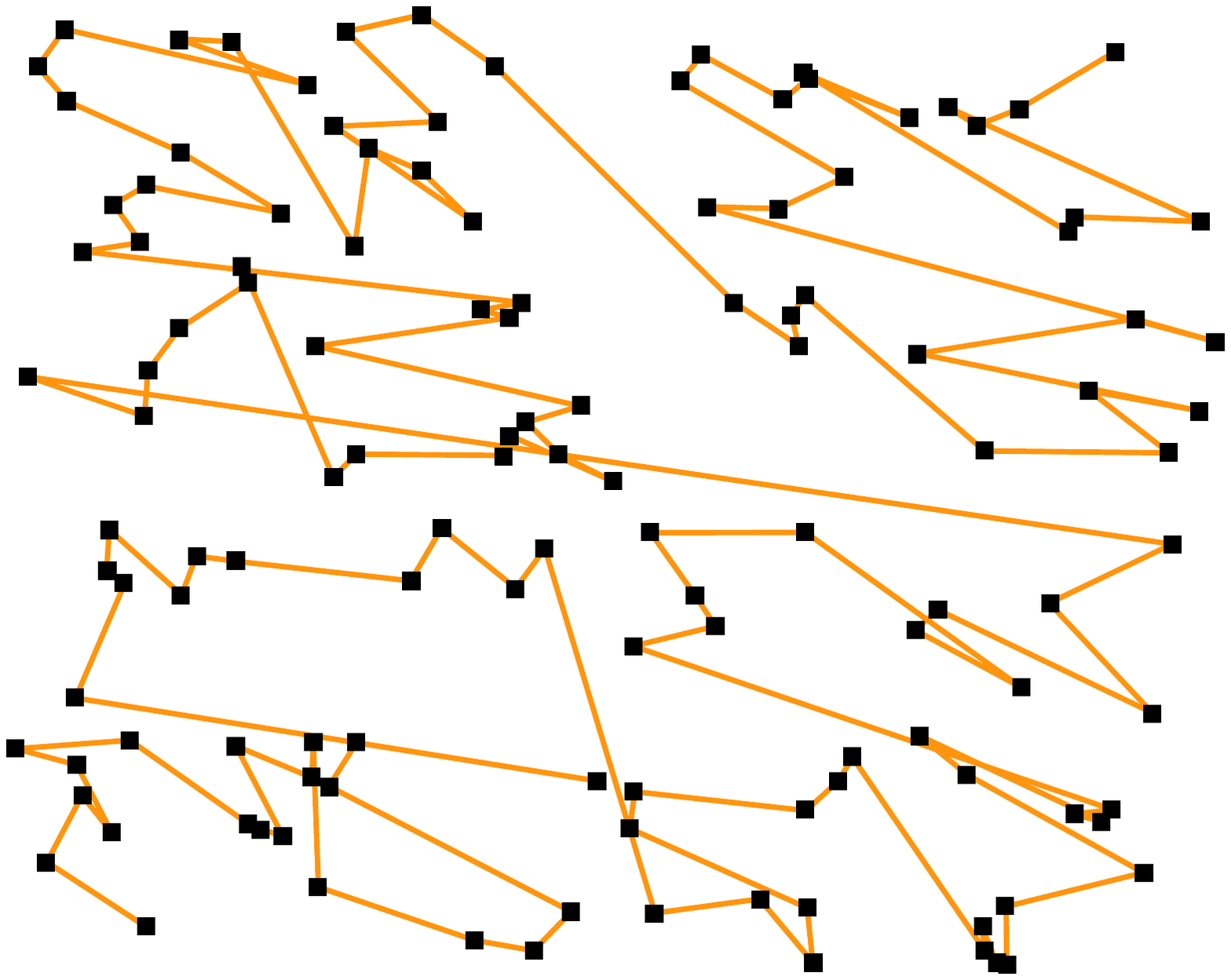}}\raisebox{5.5em}{\scalebox{0.225}[0.16875]{\includegraphics{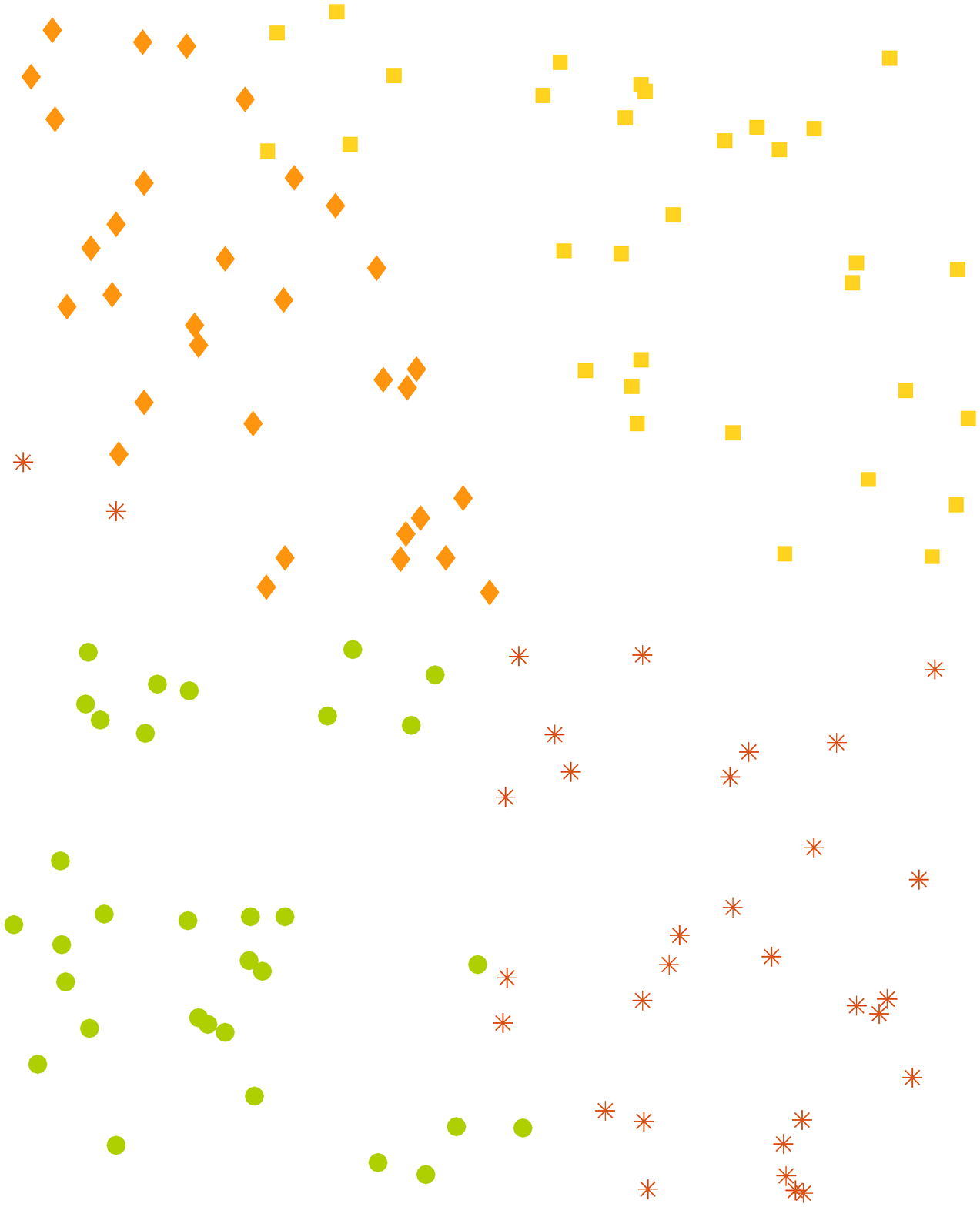}}}
\end{center}\vspace*{-6em}
\caption{\label{fig:zOrderCurve}Left: By sorting a set of arbitrary points following their Morton codes, a spatial data structure is imposed. Right: Dividing the set of ordered points in to equally sized subsets implicitly creates clusters.}
\end{figure}

We use a Z order space filling curve \cite{Morton1966} to introduce a spatial data structure on top of the input point set $\ps$. This idea is based on \cite{Lauterbach2009}. The core idea is to assign each point in $\ps$ a \textit{Morton code}, which is an integer value. By ordering the elements of $\ps$ following their Morton codes, two consecutively ordered points get spatially close to each other, cf.~Fig.~\ref{fig:zOrderCurve}. The implicit spatial structure introduced by the Morton ordering strongly simplifies the construction of the cluster tree. Whenever we have to split up a given cluster into two spatially distinct clusters in cardinality-based clustering, we only have to divide a given ordered point array into two parts, i.e.~the first halve of the elements builds the first subset and the second half of the elements builds the second subsets. That is, spatial operations get reduced to array operations.

\begin{algorithm}[t]
  \caption{Computation of Morton codes}\label{alg:mortonCodes}
  \begin{algorithmic}
    \Procedure{compute\_morton\_codes<$|\ps|$>}{coords}
	\For{each thread $t=0,\ldots, |\ps|-1$ in parallel}
		\State current\_code $\gets$ $0$
		\For{$i=1,\ldots d$}
			\State code\_current\_dim $\gets$ \Call{compute\_fixed\_point\_representation}{coords[i][$t$]}
			\State code\_current\_dim $\gets$ \Call{stretch\_bits}{code\_current\_dim, i, d}
			\State current\_code $\gets$ \Call{interleave}{code\_current\_dim,current\_code, i}
		\EndFor
		\State morton\_codes[t] $\gets$ current\_code
	\EndFor 
	\State \Return morton\_codes
    \EndProcedure
  \end{algorithmic}
\end{algorithm}

Our implementation follows the lines of \cite{Lauterbach2009}. We here assume that the reader has some knowledge about the construction of Morton codes. For details, see e.g.~\cite{MARSHALL1999}. It is trivially parallel to compute Morton codes for a point set. Algorithm~\ref{alg:mortonCodes} summarizes the corresponding parallel kernel\linebreak \textsc{compute\_morton\_codes}. Per parallel thread / point coordinate, it iterates over the dimensions of the point coordinates, where it transforms the floating-point representation of the coordinate entry to a fixed-point representation. Next, the bits of the fixed-point representation are stretched. Finally, the stretched bits are interleaved dimension-wise such that the final Morton code is constructed. Sorting the points following their Morton codes is an operation of log-linear complexity for which we assume to have an STL-like operation, cf.~Section~\ref{sec:standardizedParallelAlgorithms}.

\section{Many-core algorithms for $\Hm$ matrices}
\label{sec:manyCoreAlgorithms}
In the following, we use the beforehand discussed general parallel algorithmic patterns to construct algorithms for the many-core parallel construction of $\Hm$ matrices and the $\Hm$ matrix-vector product.

\subsection{Data structures}
\label{sec:dataStructures}
We collect the points $\ps$ in instances of a struct \texttt{point\_set}. The struct contains a multi-dimensional array \texttt{coords} of coordinates, the dimension of the points and the number of points $|\ps|$. The ordering of the point coordinates in array \texttt{coords} follows the Morton order of $\ps$, cf.~Section~\ref{sec:spatialDataStructure}. Note that, since the data structure is constructed following the Morton order while the vector $\vec{x}$ involved in the $\Hm$ matrix-vector product is stored following the original point ordering, we have to permute the vector $\vec{x}$ in the matrix-vector product or once at the beginning.

As described in Section~\ref{sec:hMatrixAlgorithms}, the $\Hm$ matrix method strongly relies on sub-blocks $\left. A_{\kernel, \ps\times \ps}\right|_{\tau\times\sigma}$ of matrix $A_{\kernel, \ps\times \ps}$, which are constructed over index blocks $\tau\times\sigma\subset I\times I$. As we will see, clusters $\tau\subset I$ will always correspond to points which are (by Morton ordering) consecutively stored in \texttt{coords}. Therefore, we can define $\tau$ by index ranges $\{i_{l, \cdot}, i_{l, \cdot}+1, i_{l,\cdot}+2, \ldots, i_{u, \cdot}\}$ pointing to the storage location in \texttt{coords}. That is, each cluster $\tau$ is represented just by the lower and upper index bounds $i_{l,\cdot}$ and $i_{u,\cdot}$.

In our implementation, we collect the nodes $w\in V_{I\times I}$ of the block cluster tree $\mathcal{T}_{I\times I}$ in instances of structs \texttt{work\_item}. In addition to the lower and upper index bounds for sets $\tau$ and $\sigma$, this struct defines storage for bounding boxes for the points corresponding to clusters $\tau$, $\sigma$ and an admissibility flag.

\subsection{Block cluster tree traversal}
The construction and traversal of the block cluster tree is based on a modified version of the tree traversal procedure presented in Algorithm~\ref{alg:treeTraversal}. Each node $w\in V_{I\times I}$ is an instance of a struct \texttt{work\_item}, cf.~Section~\ref{sec:dataStructures}. The root node is initialized to the set $I\times I$. Before computing the number of children via \textsc{compute\_child\_count}, we compute the bounding box lookup table and the map to the bounding box lookup table, cf.~Section~\ref{sec:batchedBoundingBoxComputation}. A special instance of the \textsc{compute\_child\_count} method evaluates the admissibility condition \eqref{eq:admissibilityCondition} using the precomputed bounding boxes and writes the number of children according to that result. The generic \textsc{compute\_children} method is replaced by a method that -- depending on the admissibility condition -- either creates new children by splitting up the index sets corresponding to each cluster $\tau$ or puts the node as admissible or non-admissible leave node to a parallel work queue \texttt{work\_queue} of \texttt{work\_item} structs, cf.~Section~\ref{sec:parallelOutputQueues}.

\subsection{Batched bounding box computation}
\label{sec:batchedBoundingBoxComputation}
As part of the traversal of the block cluster tree, we have to evaluate the admissibility condition~\eqref{eq:admissibilityCondition} for index blocks $\tau\times\sigma$ involving the bounding boxes of $\tau$ and $\sigma$ in each node. In the following, we will discuss an algorithm to concurrently compute the bounding boxes for clusters $\tau$, $\sigma$ in all nodes on a given level $l$ of the cluster tree. The algorithm is based on batching, cf.~Section~\ref{sec:batching}.

We collect the set of nodes on a level $l$ of the block cluster tree, i.e.~$V_{I\times I}(l)$, in the array \texttt{node\_data} of length $|V_{I\times I}(l)|$ composed of structs \texttt{work\_item} and have the input points $\ps$ in an instance of struct \texttt{point\_set}, cf.~Section~\ref{sec:spatialDataStructure}. As simplification, we only consider the concurrent computation of the bounding boxes for one cluster set, e.g.~$\tau$, in each node. 

\begin{figure}
\begin{center}
\scalebox{0.08}{\includegraphics{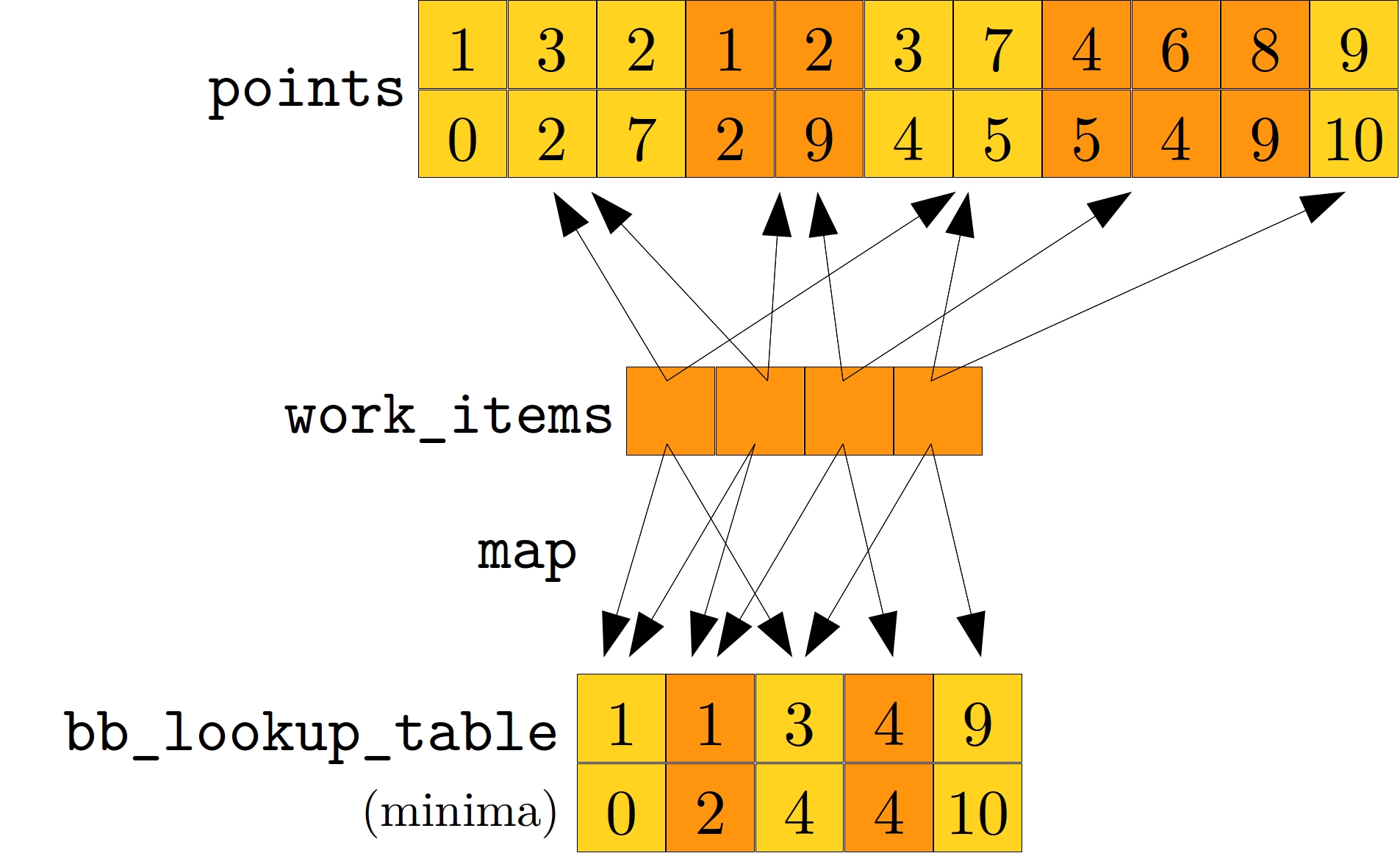}}
\end{center}
\caption{\label{fig:computeBBLookupTable}The boundary box computation is sped up by computing boundary boxes of each subset once. They are stored in \texttt{bb\_lookup\_table} and accessed via a map between work items and the lookup table.}
\end{figure}

By construction, many nodes $w\in V_{I\times I}(l)$, i.e.~on the same level of the block cluster tree, contain identical clusters (not blocks), we e.g.~have $\tau_1\times\sigma_1 = \gamma(w_1)$, $\tau_2\times\sigma_2 = \gamma(w_2)$ while $\tau_1=\tau_2$. Therefore, we first identify the set of unique clusters. We then create a lookup table \texttt{bb\_lookup\_table} storing for each unique cluster the bounding box information. In addition, we need a map from a node in \texttt{node\_data} to the entry in the lookup table. Figure~\ref{fig:computeBBLookupTable} exemplifies this idea.

\begin{algorithm}[t]
  \caption{Compute bounding box lookup table}\label{alg:computeBBLookupTable}
  \begin{algorithmic}
    \Procedure{compute\_bounding\_box\_lookup\_table}{node\_data, coords, $l$, $|V_{I\times I}(l)|$}
	\State (lower\_bounds, upper\_bounds) $\gets$ \Call{get\_index\_bounds}{node\_data}
	\State \Call{stable\_sort}{lower\_bounds}
	\State \Call{stable\_sort}{upper\_bounds}
	\State \Call{unique}{lower\_bounds, unique\_lower\_bounds}
	\State \Call{unique}{upper\_bounds, unique\_upper\_bounds}
	\State lookup\_table\_size $\gets |$lower\_bounds$|$
	\State batch\_bounds $\gets$ (unique\_lower\_bounds, unique\_upper\_bounds)
	\State \Call{sequence}{unique\_set\_indices, lookup\_table\_size, 1}
	\State \Call{init<$|\ps|$>}{batch\_keys,$0$}
	\State batch\_keys $\gets$ \Call{create\_keys}{batch\_bounds, unique\_set\_indices, $|\ps|$, lookup\_table\_size}
	\State (coord\_maxima, output\_keys) $\gets$ \Call{reduce\_by\_key}{coords,batch\_keys,maximum}
	\State (coord\_minima, output\_keys) $\gets$ \Call{reduce\_by\_key}{coords,batch\_keys,minimum}
	\State \Call{remove\_by\_key}{coord\_maxima, output\_keys, 0} \Comment{Remove invalid compute results}
	\State \Call{remove\_by\_key}{coord\_minima, output\_keys, 0}
	\State bb\_lookup\_table $\gets$ (coord\_minima, coord\_maxima)
	\State \Return bb\_lookup\_table
    \EndProcedure
  \end{algorithmic}
\end{algorithm}

Algorithm~\ref{alg:computeBBLookupTable} describes our approach to compute the entries of the lookup table \texttt{bb\_lookup\_table}. Function \textsc{compute\_bounding\_box\_lookup\_table} gets as input the coordinate array \texttt{coords} of the input point set $\ps$, the nodes $V_{I\times I}(l)$ on level $l$ in \texttt{node\_data}, and further size information. First, the lower index bounds $i_{l,1}$ and upper index bounds $i_{u,1}$ are extracted from each node and stored in arrays \texttt{lower\_index\_bounds} and \texttt{upper\_index\_bounds}. By construction, the (block) cluster tree traversal based on Z-order curves only creates clusters that do not overlap and that, for a given lower index bound, will always have the same upper bound. Therefore, we can use parallel sorting and unification methods to identify the set of unique clusters. The unique clusters are collected (by their lower and upper index bounds) in \texttt{unique\_lower\_index\_bounds} and \texttt{unique\_upper\_index\_bounds}. The final step is to compute the coordinate minima and maxima in each subset. This step follows the ideas on batching, cf.~Section~\ref{sec:batching}. The \textit{batched array} is the array of coordinates. The bounds for the batches are given by the unique lower and upper index bounds and the keys for the batches are the sequence of numbers $\{1,2,\ldots\}$. Results in the batched computation that are associated to points in $\ps$ and not being part of any subset are finally removed by removing all batched compute results associated to the key $0$. 

\begin{algorithm}[t]
  \caption{Generator for map to bounding box table}\label{alg:createMapForBB}
  \begin{algorithmic}
    \Procedure{create\_map\_for\_bounding\_boxes}{node\_data, $l$ $|V_{I\times I}(l)|$}
	\State (lower\_bounds, upper\_bounds) $\gets$ \Call{get\_index\_bounds}{node\_data}
	\State \Call{sequence<$|V_{I\times I}(l)|$>}{permutation, $|V_{I\times I}(l)|$} \Comment{permutation $\gets \{0,1,\ldots, |V_{I\times I}(l)|\}$}
	\State \Call{stable\_sort\_by\_key}{permutation, lower\_bounds}
	\State \Call{init<$|V_{I\times I}(l)|$>}{map, 0}
	\State \Call{set\_bounds\_for\_map<$|V_{I\times I}(l)|$>}{map, lower\_bounds}
	\State \Call{inclusive\_scan}{map, map, $0$, $|V_{I\times I}(l)|$} 
	\State \Call{permute\_map<$|V_{I\times I}(l)|$>}{map, permutation}
	\State \Return map
    \EndProcedure
  \end{algorithmic}
\end{algorithm}

\begin{figure}
\begin{center}
\scalebox{0.1}{\includegraphics{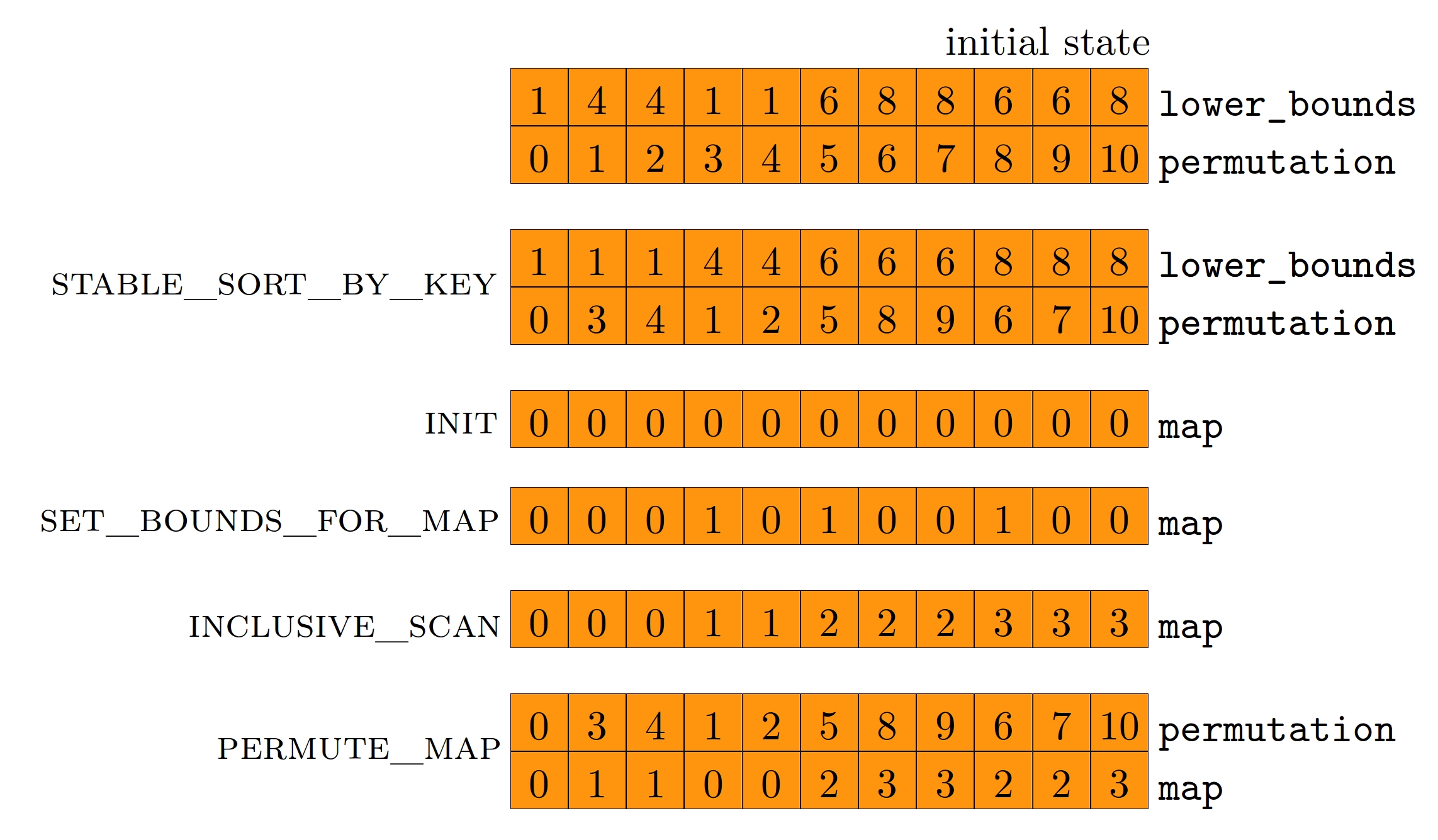}}
\end{center}
\caption{\label{fig:createMapForBB}Creating the map between a work item and an entry in the lookup table requires sorting a compute kernel and a scan operation.}
\end{figure}

Our approach to compute the map between the nodes in \texttt{node\_data} and the lookup table is summarized in Algorithm~\ref{alg:createMapForBB}. Again, we first get the lower and upper index bounds. Then, without loss of generality, we sort the lower bounds of the subsets and keep the applied permutation in \texttt{permutation}. Next, we create a global array \texttt{map} of length $|V_{I\times I}(l)|$ and initialize it to ``$0$''. The parallel kernel \texttt{set\_bound\_for\_map} of $|V_{I\times I}(l)|$ threads then sets a ``$1$'' in \texttt{map} wherever there are two different subsequent entries in the sorted \texttt{lower\_bounds}. By an inclusive scan on \texttt{map}, we create growing indices in \texttt{map} marking identical entries in \texttt{lower\_bounds}. The result is exemplified in Fig.~\ref{fig:createMapForBB}. We finally permute back \texttt{map} by kernel \texttt{permutation} with $|V_{I\times I}(l)|$ threads leading to the required map.

\subsection{Numerical linear algebra}
\label{sec:numericalLinearAlgebra}
\begin{figure}
\begin{center}
\raisebox{2.5em}{\scalebox{0.04}{\includegraphics{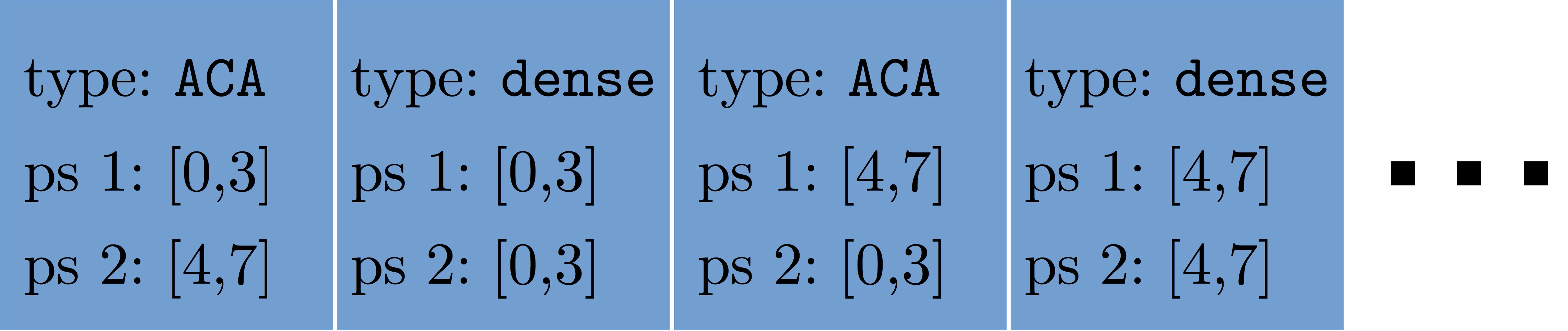}}}\quad\quad\scalebox{0.04}{\includegraphics{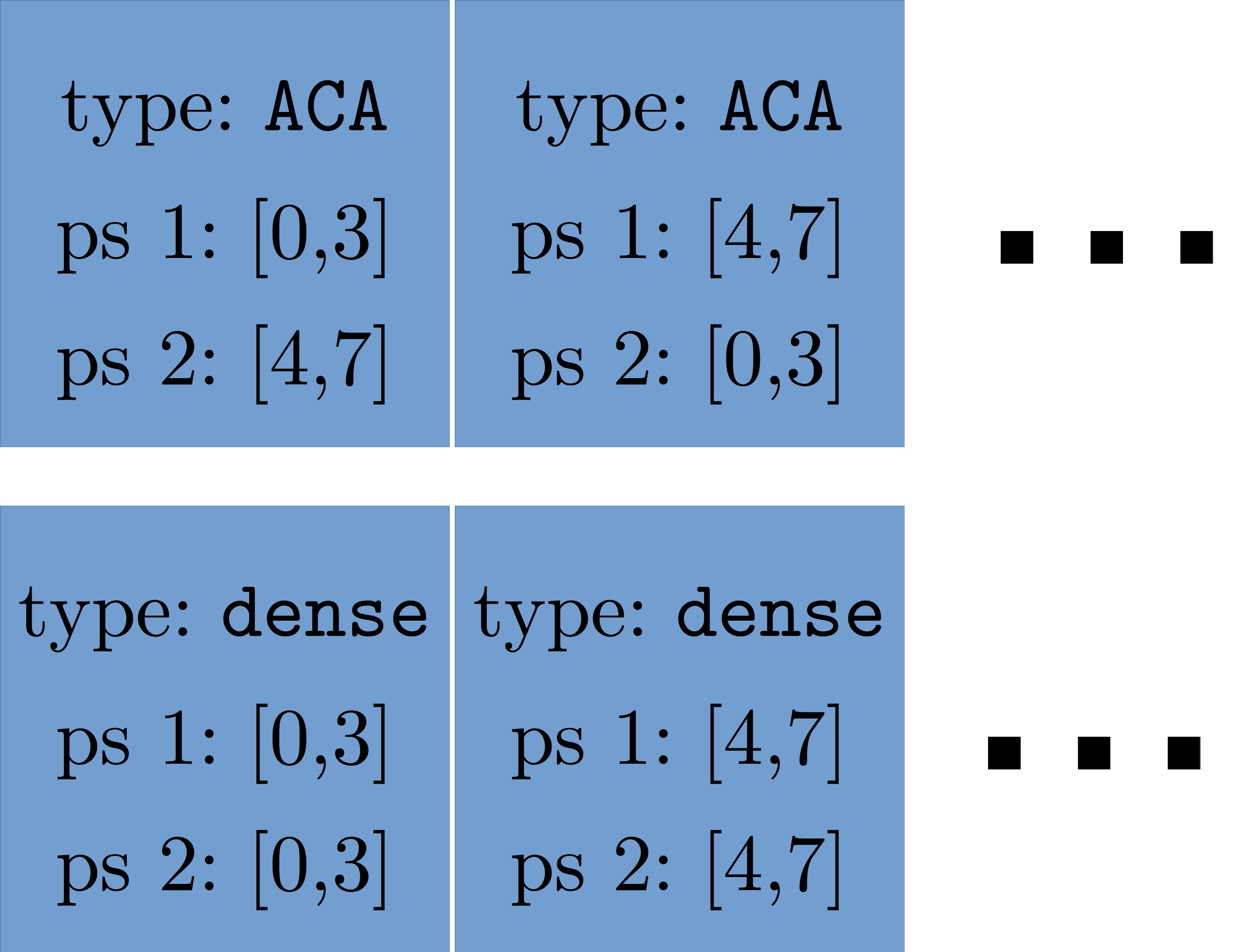}}\,
\end{center}
\caption{\label{fig:workQueue}\textit{Left:} The work queue generated during the block cluster tree traversal only contains meta information. No matrix element has been evaluated, yet. \textit{Right:} The work queue is split up into admissible and non-admissible work elements (\texttt{ACA} vs.~\texttt{dense}) before it is used as ordering for the batched linear algebra operations.}
\end{figure}

During the block cluster tree traversal, an array \texttt{work\_queue} of \texttt{work\_item} structs is constructed (via the parallel output queue), cf.~Fig.~\ref{fig:workQueue}. It contains the matrix sub-block information of blocks which are either approximated by ACA or directly constructed as dense matrices, i.e.~admissible or non-admissible. Note that we did not evaluate a single matrix entry up to this point. So we only work on meta data. 
We initially decompose the \texttt{work\_queue} into two according sub-arrays \texttt{aca\_work\_queue} and \texttt{dense\_work\_queue}, cf.~Fig.~\ref{fig:workQueue}. For the sub-matrices represented by the entries of these arrays, we either apply adaptive cross approximation or dense matrix-vector operations.

In classical (sequential) $\Hm$ matrix implementations, both, the factors $U$ and $V$ of the adaptive cross approximation \textit{and} the dense matrix blocks are precomputed during an initialization phase and then stored in memory. This is due to the fact, that often, e.g.~in boundary element methods, the evaluation of a single matrix entry is already considered very expensive, a storage operation in memory is relatively cheap and large amounts of (CPU) memory are available. Using many-core processors, this balance is somewhat different. Here, evaluating matrix elements is often much faster. However storing data in global memory, i.e.~not keeping it in the local memory of the kernel, is rather expensive. Moreover, the memory of many-core processors is often very limited. Therefore, we adapt the classical strategy to the abilities of many-core processors in the following way: We normally always re-compute all low-rank approximations and re-assemble dense matrices during each application of the fast matrix-vector product. Thereby we do not run into the very strong memory limitations of many-core processors. However, we also add the option to pre-compute the construction of the factors $U$ and $V$ in the adaptive cross approximation once, while using these factors during many matrix-vector products. Note however that this is very memory-consuming. A pre-computation of the dense sub-blocks is never done.

In the following, the details of the batched computation and application of the adaptive cross approximation and the dense matrix-vector products are presented.
 
\subsubsection{Batched adaptive cross approximation}
\begin{figure}
\begin{center}
\scalebox{0.1}{\includegraphics{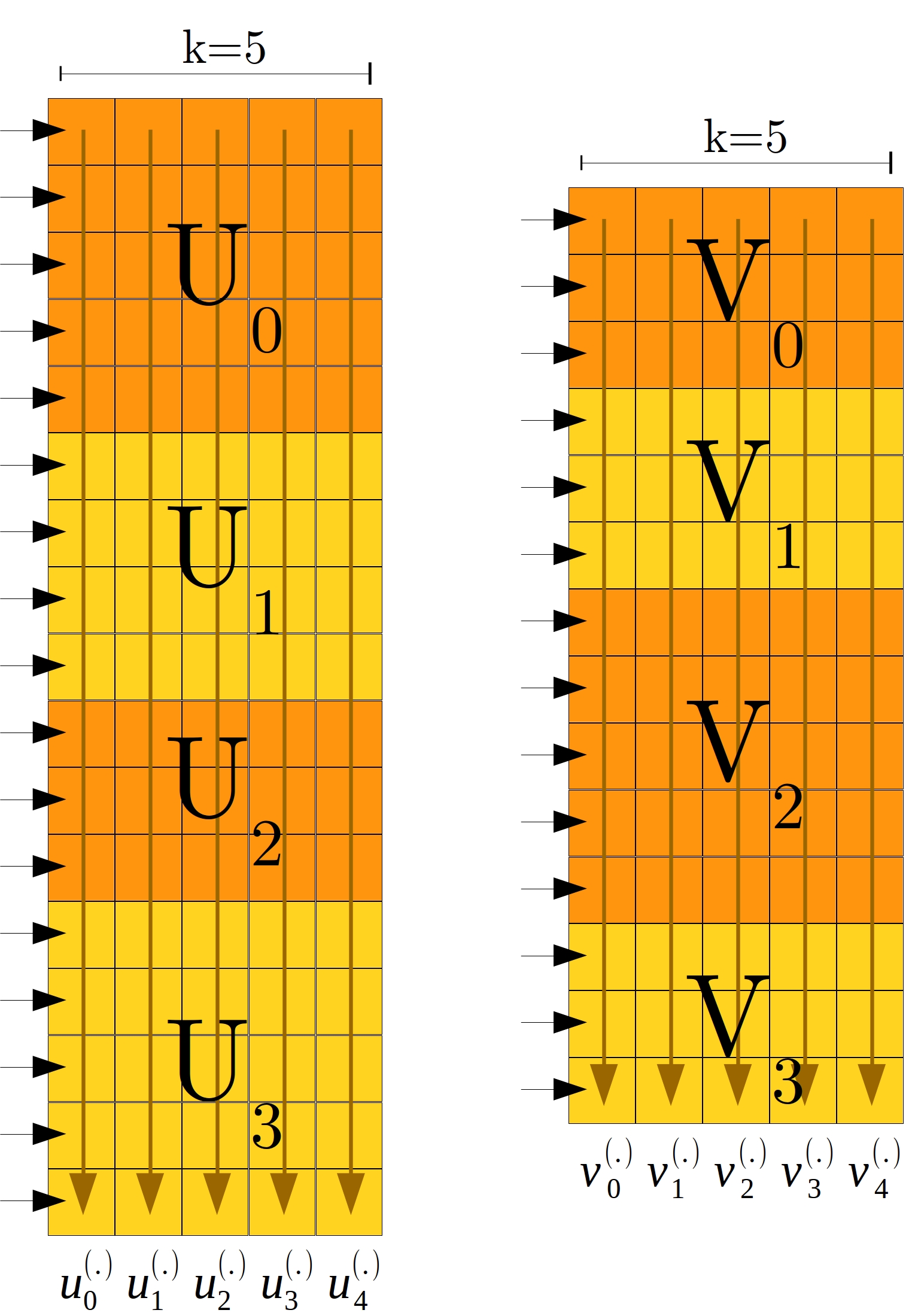}}
\end{center}
\caption{\label{fig:acaBatching}In the batched version of the adaptive cross approximation, the columns of the low-rank factors $U_l$ and $V_l$ are stored consecutively in memory. The brown arrows indicate the order in memory while the black arrows indicate threads in the parallelization.}
\end{figure}
We apply batching to compute and apply adaptive cross approximations for all ACA elements in the \texttt{aca\_work\_queue}. The storage pattern is to consecutively store elements $\vec{u}_l^{(0)}, \vec{u}_l^{(1)},\ldots \vec{u}_l^{(m-1)}$ in memory, where a single ACA sub-matrix $U^{(i)}$ is given as $U^{(i)}=(\vec{u}_1^{(i)} \vec{u}_2^{(i)} \ldots \vec{u}_k^{(i)})$. The top index is the batch number and $l$ is the index of the rank-one information. The blocks of batched rank-one information is then stored consecutively for $l=0,\ldots, k-1$, where $k$ is the maximum number of ranks that is initially given as user argument. Figure~\ref{fig:acaBatching} shows this storage principle.

In the batched ACA computation, we first set up several meta data arrays describing mainly mappings between the batched ACA data, indices of the input point set and the work items in the \texttt{aca\_work\_queue}. These mappings are used to have constant-time access in kernels being parallelized over the points, over the \texttt{aca\_work\_queue} entries or over the batched ACA data. We can compute these maps similar to the approaches presented e.g.~in Algorithm~\ref{alg:batchingKeys}. Then, we execute the classical ACA algorithm in a batched version. That is, simple transformations can be applied directly to the full batched array while batch-wise reductions are handled as described in Section~\ref{sec:batching}. 

Note that the ACA algorithm has an data-dependent iterative behavior, e.g.~in case of pivoting. That is, the algorithm might need different numbers of iterations for different batches. We cope with this by introducing a voting mechanism which stops iterations on batched data, whenever all batches have finished their work. A drawback of this approach is that the runtime for the batched version is bound from above by the \textit{slowest} batch. However, from our practical tests, this has never been a performance issue.

Depending on the choice of pre-computing or directly applying the low-rank factors $U$ and $V$, we either keep these factors in global memory for later use or we directly apply them using BLAS library calls for dense matrix-vector products. 

If we choose to recompute the ACA during each matrix-vector product, we further have the opportunity to split up the whole batched ACA computation to several smaller batched ACA operations. This allows to approach much larger matrices, which would otherwise not fit into GPU memory. To make this possible, we have to choose the number $m$ of matrix batches per batched matrix. We have designed a heuristics, which fills up a batched matrix with matrices of size $n_{b_i}\times k$ as long as $\sum_i n_{b_i}$ is smaller than a threshold $bs_{ACA}$, i.e.~the batching size for ACA. As we will see in Section~\ref{sec:batchingSizeResults}, the choice of this batching size parameter is important for the performance of the code.

\subsubsection{Batched dense sub-matrix application}
The application of the dense sub-matrix matrix-vector products is also done in a batched, parallel way. Analogously to the batched ACA computation, we first assemble, entirely in parallel, a larger number of dense sub-blocks using an appropriate compute kernel. The storage principle is similar to the one presented in the previous paragraph, i.e.~we stack the dense matrices of size $n_{b_i}\times n_{b_i}^\prime$ on top of each other. To get a simpler representation in memory, we pad all batched sub-blocks by zero columns such that they have all the same column count $\max_i n_{b_i}^\prime$. Afterwards, we use a batched version of \textit{BLAS} for the dense matrix-vector products.

As in the case of batched ACA computation, we have designed a heuristics to create batches of fixed maximum size. In case of the batched dense matrix-vector products, we choose to keep the total batch storage size smaller than a threshold $bs_{dense}$, 
$$ \max_i n_{b_i}^\prime\cdot \sum_i n_{b_i} \leq  bs_{dense}\,.$$

\section{Results}
\label{sec:performanceResults}
In this section, we evaluate the performance of the above described many-core parallel algorithms in the concrete GPU implementation \texttt{hmglib} \cite{Zaspel2017} by the author. The library is available via GitHub and is licensed under LGPL License Version 3.0. This implementation only covers the $\Hm$ matrix construction or setup and $\Hm$ matrix vector product for a matrix $A_{\kernel,\ps_1\times \ps_2}$ for a given kernel function $\kernel$ and sets $\ps_1$ and $\ps_2$. It is not intended to be feature-complete, i.e.~providing the full $\Hm$ matrix algebra. Instead, it is a test bead for the above discussed many-core parallel algorithms. Nevertheless, it is possible to solve linear systems of type \eqref{eq:denseLinearSystem} by using the iterative dense linear solvers library \texttt{MPLA} \cite{Zaspel2017a} by the author (open source, available on GitHub), which has an interface to \texttt{hmglib}. However, the objective of this benchmark chapter is to stick to the discussion of the construction and the $\Hm$ matrix-vector products, avoiding to confuse the reader by solver details and with two different library implementations.

In the following, we start our discussion by giving brief details on the library \texttt{hmglib} with the targeted hardware and applied external many-core parallel libraries. Afterwards, we introduce a model problem and show empirically that the implemented approximate matrix-vector product converges exponentially in the number ranks used in the adaptive cross approximation for the given model problem. Since the main goal is, to get a code of optimal complexity, we check the runtime complexity of \texttt{hmglib} by numerical experiments. Thereafter, we give details about the performance improvements made by batching. In fact, these performance improvements are the most relevant ones for our final results.  We finish this section, by comparing the runtimes of \texttt{hmglib} against a reference CPU implementation. Note here, that we will compare a \textit{sequentially} used multi-purpose state-of-the-art, open source library for hierarchical matrices (\texttt{H2Lib} \cite{Boerm2017}) with a very specific, parallel many-core implementation. This comparison is non-optimal. Therefore, the results of this study are only treated as a rough hint towards the actual performance improvement by using \texttt{hmglib}.

\subsection{GPU implementation \texttt{hmglib}}
The library \texttt{hmglib} \cite{Zaspel2017} is implemented for graphics processing units of\linebreak \textit{NVIDIA Corporation}. Our notion of a compute kernel from Section~\ref{sec:kernels} can be easily mapped to the compute kernels in the C language extension \textit{CUDA} for programming \textit{NVIDIA} GPUs. Note however, that an implementation in \textit{OpenCL} (for \textit{NVIDIA} and \textit{AMD} GPUs) or \textit{OpenMP} with extensions for \textit{Intel Xeon Phi} devices should be equally simple. Within our hand-implemented \textit{CUDA} compute kernels, we always use a so-called \textit{block size} of 512, i.e.~512 threads are bundled in a block with common shared memory (which we actually do not explicitly use). \texttt{hmglib} uses the \textit{CUDA Toolkit 8.0}. It is compiled with optimization parameter \texttt{-O3}. As CPU code compiler, \texttt{gcc} 4.8.5 is used.

Within our many-core parallel algorithms in Section~\ref{sec:manyCoreAlgorithms}, we launch, beside of compute kernels, library calls for general many-core parallel STL-type algorithms. In \texttt{hmglib}, the library \texttt{Thrust}, which is delivered as part of the \textit{CUDA Toolkit}, provides these STL-type algorithms. \texttt{Thrust} contains all the necessary parallel algorithms and delivers decent performance for GPUs. Moreover, we use BLAS-type operations of the library \texttt{CUBLAS}, which is also delivered as part of the \textit{CUDA Toolkit}. In case of the batched application of dense matrix-vector products, we apply the state-of-the-art GPU \textit{Lapack} library \texttt{Magma} 2.2.0. There, we specifically use the batched multiplication\linebreak \texttt{magmablas\_dgemv\_vbatched}.

\texttt{hmglib} allows to select, whether batching is applied in the matrix-vector product, or not. Moreover, it is possible to switch on the pre-computaion of the low-rank factors in the adaptive cross approximation. This requires a lot of GPU memory. However, $\Hm$ matrix-vector products can be applied faster if the low-rank factors do not have to be recomputed for each multiplication. Remember that in CPU-based $\Hm$ matrix implementations, the dense sub-blocks of the approximated matrix are often pre-computed, too. This is not done here, due to limited GPU memory and very fast matrix assembly on GPU. All calculations are done in double precision.

\subsection{Model problem}
All benchmarks consider matrix-vector products of the form
\begin{equation*}A_{\kernel, \ps\times\ps} \vec{x} \end{equation*}
with
\begin{equation*} A_{\kernel,\ps\times \ps} = \left(
   \begin{array}{ccc}
     \kernel (\vec{y}_1,\vec{y}_1) & \cdots & \kernel (\vec{y}_1,\vec{y}_{N}) \\
     \vdots & \ddots & \vdots \\
     \kernel (\vec{y}_{N},\vec{y}_1) & \cdots &
     \kernel (\vec{y}_{N},\vec{y}_{N})
   \end{array}
\right)\,,\quad \vec{x}\in\R^{N}\,.
\end{equation*}
where $\ps := \left\{\vec{y}_1,\ldots, \vec{y}_{N}\right\} \subset \Omega$ is a set of $N$ points in a space $\Omega\subset \R^d$ and $\kernel : \Omega\times\Omega\rightarrow \R$ is a bivariate {kernel} function operating on that domain. We specifically choose $\Omega = [0,1]^d$ with $d=2,3$. Moreover, the point set is a Halton sequence, i.e.~a quasi Monte-Carlo sequence, of length $N$ in $d$ dimensions. This choice corresponds to the typical setup in kernel-based approximation on the unit square / cube. We test the implementation with different (unscaled) kernel functions, namely the Gaussian kernel
\begin{equation*}
\kernel_G (\vec{y},\vec{y}^\prime) = e^{-\|\vec{y} - \vec{y}^\prime\|^2}
\end{equation*}
and a Mat\'ern kernel \cite[Section 4.4]{Fasshauer2007}
\begin{equation*}
\kernel_M (\vec{y},\vec{y}^\prime) = \frac{K_{\beta-\frac{d}{2}}(\|\vec{y}-\vec{y}^\prime\|)\|\vec{y}-\vec{y}^\prime\|^{\beta-\frac{d}{2}}}{2^{\beta-1}\Gamma(\beta)}\,,
\end{equation*}
where $K_\nu$ is the modified Bessel function of second kind of order $\nu$ and $\Gamma$ is the gamma function. We choose $\beta-\frac{d}{2}=1$. The resulting matrix $A_{\kernel_M,\ps\times\ps}$ shows up in first-order convergent function interpolation schemes in kernel-based interpolation \cite[Theorem 14.5, Example 15.4]{Fasshauer2007} for appropriately smooth functions. The norm $\|\cdot\|$ is the usual Euclidean norm of appropriated dimensionality.

This model represents the application fields of mesh-free kernel-based approximation, (non-regularized) kernel ridge regression and, in some cases, Gaussian process regression. 

\subsection{Hardware setup and time measurements}
While a major part of the development work has been carried out on the cluster Titan at Oak Ridge National Lab, the benchmarking was done on the \textit{PSG Cluster} of \textit{NVIDIA Corporation}. On the latter one, IBM S822LC compute nodes with \textit{IBM POWER8} architecture were used. They are each equipped with two 10-core \textit{IBM POWER8} processors at 2.86 GHz, 512 GB RAM and four NVIDIA Tesla P100 SXM2. Only one out of these four GPUs was used. Our CPU performance comparison is done on the same platform. Additionally, we give timings for a standard \textit{iMac} with \textit{Intel Core i5} processor at 3.2 GHz and 16 GB RAM.

Whenever we use GPU-based calculations, we use \textit{CUDA Events} to get very accurate time measurements. The time required by potentially necessary data transfers between GPU and CPU is always included. However, we assume the initial data, i.e.~the point set $\ps$ to reside in GPU memory. In case of CPU-based $\Hm$ matrix benchmarks, we use the \textit{gettimeofday} command to do the measurements. All measurements (GPU and CPU) are averaged results over five trials of a $\Hm$ matrix construction or a $\Hm$ matrix-vector product with different random vectors $\vec{x}$.

\subsection{Convergence of the matrix-vector product approximation}\label{sec:convergenceResults}
We start our experiments by checking the convergence of our $\Hm$ matrix implementation for growing ACA rank $k$ for all discussed kernel functions in two and three dimensions and problem size $N=32768$. Furthermore, we choose $C_{leaf}=256$ and $\eta=1.5$. All other parameters are not relevant for this convergence study. As for the performance measurements, we perform five runs and average over each result. The error in each run is the relative error
$$e_{rel} = \frac{\|\Hm(A_{\kernel,\ps\times\ps}) \vec{x}_{rand} - A_{\kernel,\ps\times\ps}\vec{x}_{rand}\|_2}{\|A_{\kernel,\ps\times\ps} \vec{x}_{rand}\|_2}$$
 for a random input vector $\vec{x}_{rand}$. $\Hm(A_{\kernel,\ps\times\ps})$ is the $\Hm$ matrix approximation of the full system matrix $A_{\kernel,\ps\times\ps}$. Note that we are strongly limited in the problem size $N$ since we do all computations on GPU and therefore have to do the full matrix vector product $A_{\kernel,\ps\times\ps}\vec{x}_{rand}$ in GPU memory. 

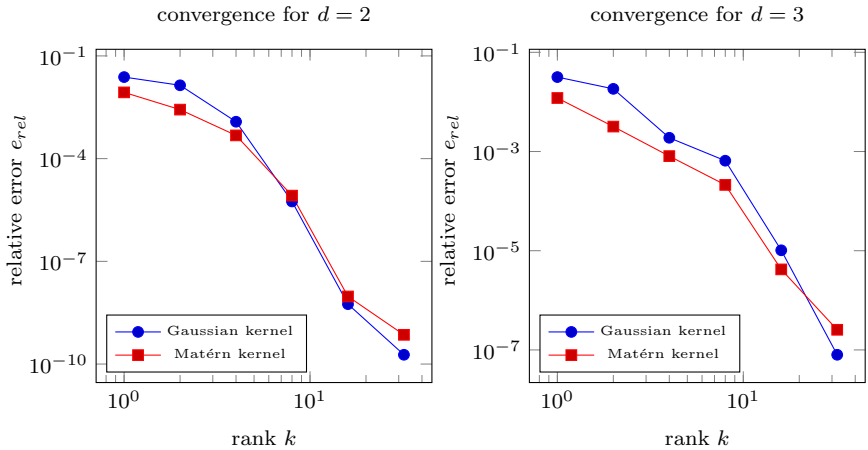
\begin{figure}
\begin{center}
\begin{tikzpicture}
\begin{loglogaxis}[height=6cm,
width= 6cm,
xlabel=rank $k$,
ylabel=relative error $e_{rel}$,
ylabel style={at={(0.04,0.55)}},
title={convergence for $d=2$},
legend style={legend pos = south west, font=\tiny}]
\addplot table{convergence_32768_dim_2_kernel_1.txt};
\addplot table{convergence_32768_dim_2_kernel_2.txt};
\legend{Gaussian kernel, Mat\'ern kernel}
\end{loglogaxis}
\end{tikzpicture}\begin{tikzpicture}
\begin{loglogaxis}[
height=6cm,
width= 6cm,
xlabel=rank $k$,
ylabel=relative error $e_{rel}$,
ylabel style={at={(0.04,0.55)}},
title={convergence for $d=3$},
legend style={legend pos = south west, font=\tiny}]
\addplot table{convergence_32768_dim_3_kernel_1.txt};
\addplot table{convergence_32768_dim_3_kernel_2.txt};
\legend{Gaussian kernel, Mat\'ern kernel}
\end{loglogaxis}
\end{tikzpicture}
\end{center}
\caption{\label{fig:convergenceResult}For fixed problem size $N=32768$ and growing number of ranks in the adaptive cross approximation, the $\Hm$ matrix-vector product converges to the full matrix-vector product with exponential convergence for dimensions $d=2$ (\textit{left}) and $d=3$ (\textit{right}).}
\end{figure}

Fig.~\ref{fig:convergenceResult} shows on the left-hand side the convergence results for $d=2$ and the two different kernels from the model problem. Our implementation delivers exponential convergence in the number $k$ of ranks used in the adaptive cross approximation. The same test is repeated for dimension $d=3$ with similar results. Since the results for Gaussian and Mat\'ern kernel are almost identical, we will, in the following paragraphs, restrict ourselves to performance studies for the Gaussian kernel.

\subsection{Runtime complexity and performance of the GPU-parallel code}
The crucial objective of an implementation of the hierarchical matrix method is to achieve the optimal runtime complexity of $O(N \log N)$ for the matrix-vector product at fixed rank $k$. However, very often, high (pre-asymptotic) runtime performance on many-core hardware is only achieved by sticking to algorithmic simplifications of worse complexity but higher performance. The following empirical study shall show that the $\Hm$ matrix implementation in \texttt{hmglib}, which is based on our many-core parallel $\Hm$ matrix algorithms from Section~\ref{sec:manyCoreAlgorithms}, actually achieves the required $O(N \log N)$ runtime complexity. To study this, we choose $\eta=1.5$, $C_{leaf}=2048$, $k=16$, $bs_{dense}=2^{27}$ and $bs_{ACA}=2^{25}$, use batching and carry out performance measurements for growing problem size $N$. 

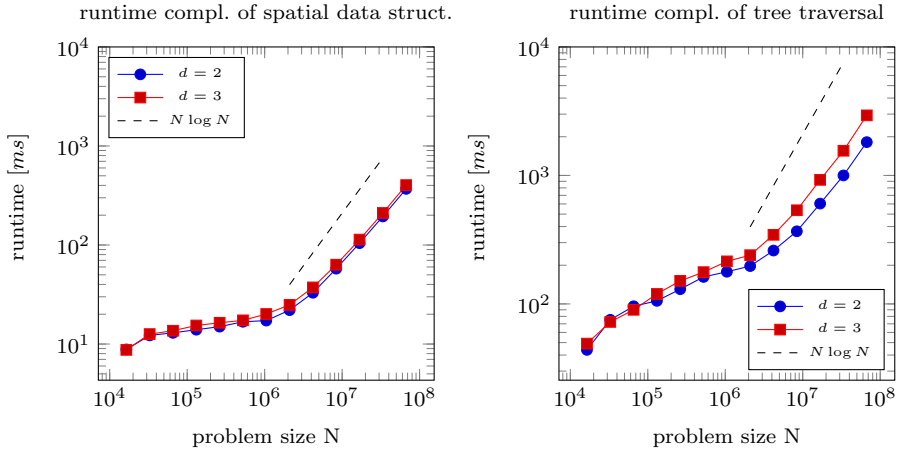
\begin{figure}
\begin{center}
\begin{tikzpicture}
\begin{loglogaxis}[height=6cm,
width= 6cm,
xlabel=problem size N,
ylabel={runtime [$ms$]},
ylabel style={at={(0.04,0.55)}},
ymax=10000,
title={runtime compl.~of spatial data struct.},
legend style={legend pos = north west, font=\tiny}]
\addplot table [x expr=2^\thisrowno{0}, y expr=\thisrowno{1}/1.0] {spatial_data_structure_benchmark_dim_2.txt};
\addplot table [x expr=2^\thisrowno{0}, y expr=\thisrowno{1}/1.0] {spatial_data_structure_benchmark_dim_3.txt};
\addplot[color=black,dashed,mark=none] expression[domain=2^21:2^25] {0.000003*x*log10(x)};
\legend{$d=2$,$d=3$, $N \log N$}
\end{loglogaxis}
\end{tikzpicture}\begin{tikzpicture}
\begin{loglogaxis}[height=6cm,
width= 6cm,
xlabel=problem size N,
ylabel={runtime [$ms$]},
ylabel style={at={(0.04,0.55)}},
ymax=10000,
title={runtime compl.~of tree traversal},
legend style={legend pos = south east, font=\tiny}]
\addplot table [x expr=2^\thisrowno{0}, y expr=\thisrowno{1}/1.0] {tree_traversal_benchmark_dim_2.txt};
\addplot table [x expr=2^\thisrowno{0}, y expr=\thisrowno{1}/1.0] {tree_traversal_benchmark_dim_3.txt};
\addplot[color=black,dashed,mark=none] expression[domain=2^21:2^25] {0.00003*x*log10(x)};
\legend{$d=2$,$d=3$,$N \log N$}
\end{loglogaxis}
\end{tikzpicture}
\end{center}
\caption{\label{fig:runtimeComplexitySetup}Even for the largest problem size of $2^{26}\approx 67$ million unknowns, the time for the spatial data structure setup is below $0.5$ seconds (\textit{left}). The tree construction and traversal requires less than 3 seconds for $2^{26}$ unknowns, while matching the required runtime complexity of $N \log N$ (\textit{right}). }
\end{figure}

We first discuss the runtime complexity of the setup of the spatial data structure. While computing the Morton codes for all points $\vec{y}_i$ is of complexity $O(N)$, sorting the points following the Z order curve is a $O(N \log N)$ operation. This is reflected by our empirical study shown on the left-hand side of Fig.~\ref{fig:runtimeComplexitySetup}. For $d=2$ and $d=3$ we observe a runtime complexity of $O(N \log N)$ after some pre-asymptotic range. The same behavior is observed for the construction and the traversal of the block cluster tree. Runtime results for this case are given on the right-hand side of Fig.~\ref{fig:runtimeComplexitySetup}. Note again that it is non-trivial to get the optimal complexity for such a complex many-core parallel code. Figure~\ref{fig:runtimeComplexitySetup} further outlines that the spatial data structure setup and the tree traversal is actually very fast. Even for $N=2^{26}$, i.e. an approximation of a dense matrix of roughly $67 \times 67$ million entries, we only need roughly $0.4$ seconds for the spatial data structure and about $3$ seconds for the tree traversal (for $d=3$).

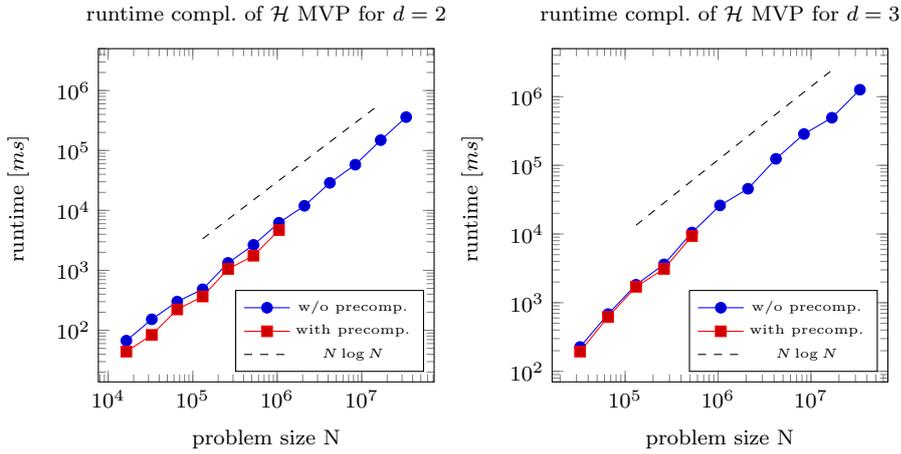
\begin{figure}
\begin{center}
\begin{tikzpicture}
\begin{loglogaxis}[height=6cm,
width= 6cm,
xlabel=problem size N,
ylabel={runtime [$ms$]},
ylabel style={at={(0.04,0.55)}},
ymax=5000000,
title={runtime compl.~of $\Hm$ MVP for $d=2$},
legend style={legend pos = south east, font=\tiny}]
\addplot table [x expr=2^\thisrowno{0}, y expr=\thisrowno{1}/1.0] {complexity_benchmark_without_precomputing_dim_2.txt};
\addplot table [x expr=2^\thisrowno{0}, y expr=\thisrowno{1}/1.0] {complexity_benchmark_with_precomputing_dim_2.txt};
\addplot[color=black,dashed,mark=none] expression[domain=2^17:2^24] {0.005*x*log10(x)};
\legend{w/o precomp.,with precomp., $N \log N$}
\end{loglogaxis}
\end{tikzpicture}\begin{tikzpicture}
\begin{loglogaxis}[height=6cm,
width= 6cm,
xlabel=problem size N,
ylabel={runtime [$ms$]},
ylabel style={at={(0.04,0.55)}},
title={runtime compl.~of $\Hm$ MVP for $d=3$},
ymax=5000000,
legend style={legend pos = south east, font=\tiny}]
\addplot table [x expr=2^\thisrowno{0}, y expr=\thisrowno{1}/1.0] {complexity_benchmark_without_precomputing_dim_3.txt};
\addplot table [x expr=2^\thisrowno{0}, y expr=\thisrowno{1}/1.0] {complexity_benchmark_with_precomputing_dim_3.txt};
\addplot[color=black,dashed,mark=none] expression[domain=2^17:2^24] {0.02*x*log10(x)};
\legend{w/o precomp., with precomp.,$N \log N$}
\end{loglogaxis}
\end{tikzpicture}
\end{center}
\caption{\label{fig:runtimeComplexityMVP}The $\Hm$ matrix-vector product shows the optimal algorithmic complexity of $O(N \log N)$. The operation is slightly more expensive if used on points in tree dimension (\textit{right}) in contrast to points in two dimensions (\textit{left}). Using pre-computation for the ACA factors leads to a performance improvement.}
\end{figure}

The second part of this runtime complexity study covers the application of the fast matrix-vector product. Figure~\ref{fig:runtimeComplexityMVP} shows the measurements of the runtime for growing problem size $N$ and different dimensionality $d$. Within each performance plot, we further distinguish between measurements that were done using a matrix-vector product with precomputed ACA factors and with on-the-fly computation of the ACA factors. Pre-computing the ACA factors results in a performance improvement, which will be discussed in more detail in Section~\ref{sec:cpuPerformanceComparison}. In the plot, the impact is not clearly visible due to the logarithmic scaling of the axis. We cannot show runtime results with pre-computing for problem sizes beyond $N=2^{19}$ or $N=2^{20}$ due to the limited GPU memory. 

In all cases, we observe a runtime complexity of $O(N \log N)$. Moreover, even for a problem size of $N=2^{25}$, i.e.~an approximated matrix-vector product for a dense matrix of $33 \times 33$ million entries, we see a runtime of only 6 minutes for a matrix-vector product on points in two dimensions.

\subsection{Performance analysis of batching}\label{sec:batchingPerformance}
Beforehand, we discussed the performance results of our implementation using batching in all linear algebra operations, as discussed in Section~\ref{sec:numericalLinearAlgebra}. However, it is important to know that batching is one of the crucial ingredients of this code allowing for high performance of the overall method. To show the actual impact of batching, we will analyse the performance with and without batching in the linear algebra operations. However, before we come to this point, we want to address the topic of parameter choice of the batching sizes $bs_{dense}$ and $bs_{ACA}$.

\subsubsection{Batching size influence}\label{sec:batchingSizeResults}
 In Section~\ref{sec:numericalLinearAlgebra}, we introduced the parameters $bs_{dense}$ and $bs_{ACA}$ as batching sizes for the batching of the dense matrix-vector products and the batching of the adaptive cross approximation. These parameters balance the memory consumption against the performance improvement. To understand this relationship further, we benchmark the runtime of the batched dense matrix-vector products and the batched ACA in the $\Hm$ matrix-vector product for different batching sizes. It is done for $N=2^{20}$, $k=16$, $\eta=1.5$ and $d=2$. We consider results for $C_{leaf}=1024$ and $C_{leaf}=2048$.

\begin{figure}
\begin{center}
\begin{tikzpicture}
\begin{loglogaxis}[height=6cm,
width= 6cm,
xlabel=batching size $bs_{dense}$,
ylabel={runtime [$s$]},
ylabel style={at={(0.04,0.55)}},
legend style={legend pos = north east, font=\tiny},
	  title={runtime of batched dense MVP}
]
\addplot table [x expr=2^\thisrowno{0}, y expr=\thisrowno{1}/1000.0] {batching_changing_benchmark_dense_with_batching_cleaf_1024.txt};
\addplot table [x expr=2^\thisrowno{0}, y expr=\thisrowno{1}/1000.0] {batching_changing_benchmark_dense_with_batching_cleaf_2048.txt};
\legend{$C_{leaf}=1024$,$C_{leaf}=2048$}
\end{loglogaxis}
\end{tikzpicture}
\begin{tikzpicture}
\begin{loglogaxis}[height=6cm,
width= 6cm,
xlabel=batching size $bs_{ACA}$,
ylabel={runtime [$s$]},
ylabel style={at={(0.04,0.55)}},
title={runtime of batched ACA},
legend style={legend pos = north east, font=\tiny}]
\addplot table [x expr=2^\thisrowno{0}, y expr=\thisrowno{1}/1000.0] {batching_changing_benchmark_aca_with_batching_cleaf_1024.txt};
\addplot table [x expr=2^\thisrowno{0}, y expr=\thisrowno{1}/1000.0] {batching_changing_benchmark_aca_with_batching_cleaf_2048.txt};
\legend{$C_{leaf}=1024$,$C_{leaf}=2048$}
\end{loglogaxis}
\end{tikzpicture}
\end{center}
\caption{\label{fig:batchingParameters}The performance of batching strongly depends on the size of the batched array or matrices that are used. This is clearly visible for the batching of dense matrix-vector products (\textit{left}) and adaptive cross approximation (\textit{right}). The optimal batching size is only slightly influenced by the choice of the parameter $C_{leaf}$.} 
\end{figure}
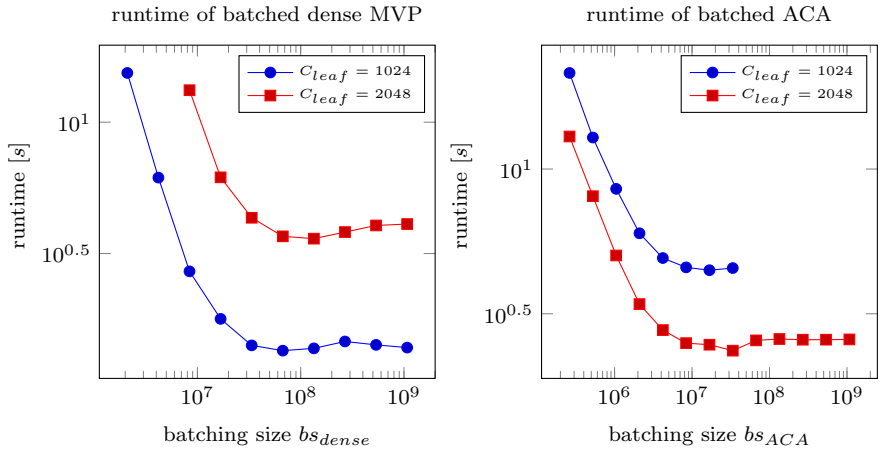

Figure~\ref{fig:batchingParameters} collects the results for the parameter studies in the batching size for the batched dense matrix-vector products on the left-hand side and for the batched ACA computation on the right-hand side. The choice of the leaf size $C_{leaf}$ has a considerable influence on the performance balance between dense matrix-vector products and ACA. That is, larger leaf sizes lead to larger runtimes in the dense matrix-vector products. However the ACA runtime is reduced. The opposite holds for smaller leaf sizes. Moreover, the smaller leaf size of $C_{leaf}=1024$ in batched ACA leads to a higher memory consumption for the batching, which limits us to a maximum of $bs_{ACA}=2^{25}$.

The general tendency in the results in Fig.~\ref{fig:batchingParameters} is that increasing the batching size increases the performance to an optimum. Beyond this optimum, the performance of the batching gets slightly worse. This performance improvement up to an optimum is due to the improvement of the occupancy of the GPU. That is, the GPU gets more work to do. Thereby, it can hide latencies etc.~\textit{behind} parallel work. The slight performance degradation beyond the optimum for larger batching sizes is maybe due to a slight over-subscription of the GPU: The maximum throughput limit is hit, however, due to more batches per batched operation, the data structure overhead becomes visible. Note, however, that this latter reasoning is speculative.

Overall, choosing an appropriate batching size is rather simple. The rule of thumb is to take it as large as possible (in terms of memory consumption) and to accept the slight performance reduction for a too large batch size.

\subsubsection{Performance improvement by batching}
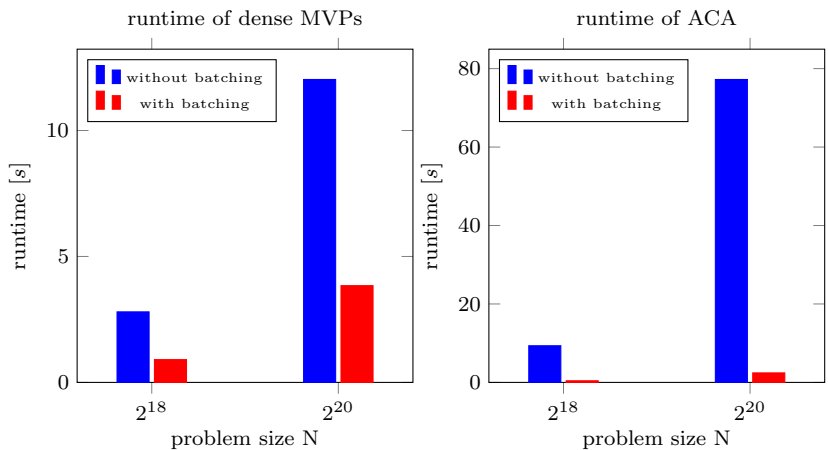
\begin{figure}
\begin{center}
\begin{tikzpicture}
        \begin{semilogxaxis}[
          ybar,
          bar width=12pt,
          height=6cm,
          width= 6cm,
          ymin=0,
          xmin=0.0,
          xtick=data, 
xticklabels={$2^{18}$,$2^{20}$},
xticklabel style={anchor=north},
enlarge x limits=0.4,
	  xlabel={problem size N},
          ylabel={runtime [$s$]},
          ylabel near ticks,
	  title={runtime of dense MVPs},
          legend style={legend pos = north west, font=\tiny}
          ]
          \addplot table [x expr=2^\thisrowno{0}, y expr=\thisrowno{1}/1000.0] {batching_benchmark_dense_without_batching_cleaf_2048.txt};
          \addlegendentry{without batching};
          \addplot table [x expr=2^\thisrowno{0}, y expr=\thisrowno{1}/1000.0] {batching_benchmark_dense_with_batching_cleaf_2048.txt};
          \addlegendentry{with batching};

        \end{semilogxaxis}
\end{tikzpicture}\begin{tikzpicture}
        \begin{semilogxaxis}[
          ybar,
          bar width=12pt,
          height=6cm,
          width= 6cm,
          ymin=0,
          xmin=0.0,
          xtick=data, 
xticklabels={$2^{18}$,$2^{20}$},
xticklabel style={anchor=north},
enlarge x limits=0.4,
          xticklabel style={anchor=north},
	  xlabel={problem size N},
          ylabel={runtime [$s$]},
          ylabel near ticks,
	  title={runtime of ACA},
          legend style={legend pos = north west, font=\tiny}
          ]
          \addplot table [x expr=2^\thisrowno{0}, y expr=\thisrowno{1}/1000.0] {batching_benchmark_aca_without_batching_cleaf_2048.txt};
          \addlegendentry{without batching};
          \addplot table [x expr=2^\thisrowno{0}, y expr=\thisrowno{1}/1000.0] {batching_benchmark_aca_with_batching_cleaf_2048.txt};
          \addlegendentry{with batching};

        \end{semilogxaxis}
\end{tikzpicture}
\end{center}
\caption{\label{fig:batchingPerformanceImprovement}We observe a significant performance improvement by roughly a factor of 32, when using batching in the ACA computation (\textit{right}). Batching dense matrix-vector products still improves performance by roughly a factor of three (\textit{left}).}
\end{figure}

We next discuss the performance improvement for batched dense matrix-vector products and for batched ACA. We use parameters $N=2^{20}$, $k=16$, $\eta=1.5$, $d=2$, $C_{leaf}=2048$, $bs_{dense}=2^{27}$ and $bs_{ACA}=2^{25}$. Figure~\ref{fig:batchingPerformanceImprovement} summarizes the results of this study with results for the batching of dense matrix-vector products on the left-hand side and results for the batched adaptive cross approximation on the right-hand side. For a problem size of $N=2^{20}$, the batched application of the dense matrix-vector products is by more than a factor of 3 faster. We do not gain more, since, for $C_{leaf}=2048$, we have a lot of large dense matrix sub-blocks which very soon fully occupy the GPU.

In contrast, the performance improvement for the adaptive cross approximation is about a factor of 32 for $N=2^{20}$. This strong impact is due to the small amount of work that is done for each individual ACA computation and is a significant contribution of this work. 

To summarize, an efficient $\Hm$ matrix-vector product would not be possible without ACA batching. However, it also pays off to do batching for the dense matrix-vector products.

\subsection{Performance comparison against \texttt{H2Lib}}
\label{sec:cpuPerformanceComparison}
In the following, we aim at relating the performance of \texttt{hmglib} to the CPU $\Hm$ and $\Hm^2$ matrix library \texttt{H2Lib} \cite{Boerm2017} in the, at time of writing this paper, latest available version. We have chosen \texttt{H2Lib}, since it is under active development and an Open Source library. The \texttt{H2Lib} library implements an algebra for $\Hm$ matrices and $\Hm^2$ matrices. That is, the library allows to construct, add, multiply, factorize, etc. $\Hm$ and $\Hm^2$ matrices. Moreover, it contains modules for the solution of problems discretized by the boundary element method. Recently, support for a GPU-accelerated $\Hm^2$ matrix setup for boundary element method problems was added \cite{Boerm2015}, as discussed in Section~\ref{sec:introduction}. \texttt{H2Lib} also contains some support for shared-memory parallelism. However, it seemed to have no impact on the performance of the $\Hm$ matrix construction and the $\Hm$ matrix-vector product. Therefore, we used the sequential version, only.

As argued before, the comparison of our GPU implementation, which only implements the $\Hm$ matrix-vector product, with this feature-complete sequential CPU implementation, which has been specifically optimized for $\Hm^2$ matrices and boundary element method problems, is non-optimal by construction. However, we add this comparison to somehow relate our performance results to currently available software in the field.  

In our performance benchmarks, we try our best to fit the \texttt{H2Lib} implementation to our GPU implementation, even if this means that we have to extend the \texttt{H2Lib} for this. To give an example, we added the ability to do ACA for a fixed rank $k$, which was not available in the library, before. On the IBM POWER8 platform, \texttt{H2Lib} is compiled with \texttt{gcc} 4.8.5 and the usual optimizations and linked against the, at time of writing this article, latest available version of \texttt{OpenBLAS}. On the Intel architecture, it is compiled with the same compiler, however linked against the default \textit{LAPACK} implementation of \textit{macOS Sierra 10.12}.

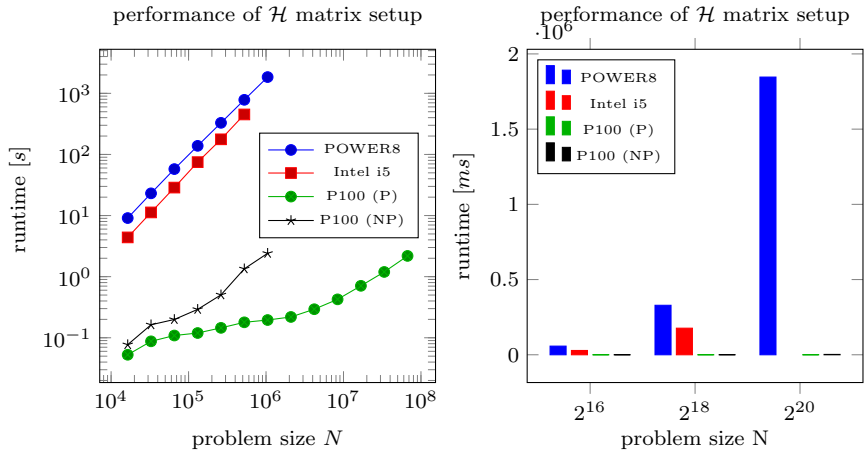
\begin{figure}
\begin{center}
\begin{tikzpicture}
\begin{loglogaxis}[height=6cm,
width= 6cm,
xlabel=problem size $N$,
ylabel={runtime [$s$]},
ylabel style={at={(0.04,0.55)}},
title={performance of $\Hm$ matrix setup},
legend style={at={(0.95,0.75)}, font=\tiny}
]
          \addplot table [x expr=2^\thisrowno{0}, y expr=\thisrowno{1}/1000.0] {cpu_power8_h2lib_setup_benchmark_dim_2.txt};
          \addlegendentry{POWER8};
          \addplot table [x expr=2^\thisrowno{0}, y expr=\thisrowno{1}/1000.0] {cpu_intel_h2lib_setup_benchmark_dim_2.txt};
          \addlegendentry{Intel i5};
          \addplot table [x expr=2^\thisrowno{0}, y expr=\thisrowno{1}/1000.0] {hmglib_setup_without_precomputing_dim_2.txt};
          \addlegendentry{P100 (P)};
          \addplot table [x expr=2^\thisrowno{0}, y expr=\thisrowno{1}/1000.0] {hmglib_setup_with_precomputing_dim_2.txt};
          \addlegendentry{P100 (NP)};

\end{loglogaxis}
\end{tikzpicture}\begin{tikzpicture}
        \begin{semilogxaxis}[
          ybar,
          bar width=6pt,
          height=6cm,
          width= 6cm,
          xmin=0.0,
          xtick=data, 
          xticklabel style={anchor=north},
	  xlabel={problem size N},
          ylabel={runtime [$ms$]},
          ylabel near ticks,
title={performance of $\Hm$ matrix setup},
          legend style={legend pos=north west, font=\tiny},
	  skip coords between index={0}{2},
	  skip coords between index={3}{4},
	  skip coords between index={5}{6},
	  skip coords between index={7}{13},
	  xticklabels={$2^{16}$,$2^{18}$,$2^{20}$},
          xticklabel style={anchor=north},
	  enlarge x limits=0.3
]
          \addplot table [x expr=2^\thisrowno{0}, y expr=\thisrowno{1}/1.0] {cpu_power8_h2lib_setup_benchmark_dim_2.txt};
          \addlegendentry{POWER8};
          \addplot table [x expr=2^\thisrowno{0}, y expr=\thisrowno{1}/1.0] {cpu_intel_h2lib_setup_benchmark_dim_2.txt};
          \addlegendentry{Intel i5};
          \addplot table [x expr=2^\thisrowno{0}, y expr=\thisrowno{1}/1.0] {hmglib_setup_without_precomputing_dim_2.txt};
          \addlegendentry{P100 (P)};
          \addplot table [x expr=2^\thisrowno{0}, y expr=\thisrowno{1}/1.0] {hmglib_setup_with_precomputing_dim_2.txt};
          \addlegendentry{P100 (NP)};
        \end{semilogxaxis}
\end{tikzpicture}
\end{center}
\caption{\label{fig:comparisonSetupPhase}We compare the runtime of the $\Hm$ matrix setup in the \textit{H2Lib} (including the computation of the ACA and all dense sub-blocks) with the setup in the \texttt{hmglib} library with \textit{(P)} and without \textit{(NP)} pre-computing the ACA. The GPU-based implementation outperforms the \textit{sequential} CPU-based implementation by more than two orders of magnitude.}
\end{figure}

We start the comparison with a benchmark of the $\Hm$ matrix construction or setup phase. In case of the \texttt{H2Lib} this construction phase contains the spatial data structure setup, the block cluster tree traversal, the pre-computation of all low-rank factors and the assembly of all dense sub-blocks of the $\Hm$ matrix. We choose $\eta=1.5$ and $C_{leaf}=128$. On the other hand, we choose $\eta=1.5$, $C_{leaf}=2048$, $bs_{dense}=2^{27}$ and $bs_{ACA}=2^{25}$ in the GPU implementation and analyse the construction phase including pre-computation \textit{(P)} of the ACA factors or without \textit{(NP)} such a pre-computation. Note that the leaf size $C_{leaf}$ has a significant impact on the performance on the method. Therefore, we adapt it for the different architectures for best possible performance. All results in this paragraph are computed for a fixed rank of $k=16$ and dimension $d=2$. Since we observed very strong fluctuations of the runtime on the Intel workstation, we always take the smallest runtime out of five trials on that architecture to be as fair as possible.

Figure~\ref{fig:comparisonSetupPhase} gives the result for the first comparison. On the left-hand side, runtimes of the setup phase are given for growing problem size. The diagram on the right-hand side directly compares the results on the different architectures for fixed problem sizes. Due to limited memory, the benchmark is stopped for $N=2^{19}$ on the Intel machine and for $N=2^{20}$ on the GPU with pre-computing. The benchmark on the POWER8 CPU system is stopped for $N=2^{20}$ due to large runtime. In case of the largest common problem size, i.e.~$N=2^{19}\approx 0.5$ million points, the CPU implementation requires 782 seconds and 451 seconds on the POWER8 system and the Intel system, respectively. On the other hand, the GPU implementation only needs 1.3 seconds with precomputing and 0.8 seconds without pre-computing. That is, it is more than two orders of magnitude faster, However, note again that the setup phase on CPU also pre-computes the dense matrix sub-blocks. Moreover, we compare a sequential implementation with a strongly parallelized GPU implementation. Even more, the single-threaded performance of the POWER8 system seems to be limited, which is why we also included the Intel workstation in the benchmark. A more fair comparison would e.g.~compare a parallel CPU code on 16 CPU cores with the GPU code. However, even in this case (assuming perfect scalability on CPU), the GPU would outperform the CPU-based version by a factor of twenty.

\begin{figure}
\begin{center}
\begin{tikzpicture}
\begin{loglogaxis}[height=6cm,
width= 6cm,
xlabel=problem size $N$,
ylabel={runtime [$s$]},
ylabel style={at={(0.04,0.55)}},
  title={performance of $\Hm$ MVP},
legend style={legend pos = south east, font=\tiny}]
          \addplot table [x expr=2^\thisrowno{0}, y expr=\thisrowno{1}/1000.0] {cpu_power8_h2lib_mvp_benchmark_dim_2.txt};
          \addlegendentry{POWER8};
          \addplot table [x expr=2^\thisrowno{0}, y expr=\thisrowno{1}/1000.0] {cpu_intel_h2lib_mvp_benchmark_dim_2.txt};
          \addlegendentry{Intel i5};
          \addplot table [x expr=2^\thisrowno{0}, y expr=\thisrowno{1}/1000.0] {complexity_benchmark_with_precomputing_dim_2.txt};
          \addlegendentry{P100 (P)};
          \addplot table [x expr=2^\thisrowno{0}, y expr=\thisrowno{1}/1000.0] {complexity_benchmark_without_precomputing_dim_2.txt};
          \addlegendentry{P100 (NP)};

\end{loglogaxis}
\end{tikzpicture}\begin{tikzpicture}
        \begin{semilogxaxis}[
          ybar,
          bar width=6pt,
          height=6cm,
          width= 6cm,
          xtick=data, 
	  xmin=0,
	  xticklabels={$2^{16}$,$2^{18}$,$2^{20}$},
          xticklabel style={anchor=north},
	  enlarge x limits=0.3,
	  xlabel={problem size N},
          ylabel={runtime [$s$]},
	  ylabel style={at={(0.07,0.55)}},
          legend style={legend pos = north west, font=\tiny},
	  skip coords between index={0}{2},
	  skip coords between index={3}{4},
	  skip coords between index={5}{6},
	  skip coords between index={7}{20},
	  title={performance of $\Hm$ MVP}
          ]
          \addplot table [x expr=2^\thisrowno{0}, y expr=\thisrowno{1}/1000.0] {cpu_power8_h2lib_mvp_benchmark_dim_2.txt};
          \addlegendentry{POWER8};
          \addplot table [x expr=2^\thisrowno{0}, y expr=\thisrowno{1}/1000.0] {cpu_intel_h2lib_mvp_benchmark_dim_2.txt};
          \addlegendentry{Intel i5};
          \addplot table [x expr=2^\thisrowno{0}, y expr=\thisrowno{1}/1000.0] {complexity_benchmark_with_precomputing_dim_2.txt};
          \addlegendentry{P100 (P)};
          \addplot table [x expr=2^\thisrowno{0}, y expr=\thisrowno{1}/1000.0] {complexity_benchmark_without_precomputing_dim_2.txt};
          \addlegendentry{P100 (NP)};

        \end{semilogxaxis}
\end{tikzpicture}
\end{center}
\caption{\label{sec:comparisonMVP}The GPU-based $\Hm$ matrix-vector product outperforms the \textit{single}-threaded CPU-based $\Hm$ matrix vector product by one order of magnitude. By precomputing \textit{(P)} the ACA factors we get an increase in performance by about 60 \%.}
\end{figure}
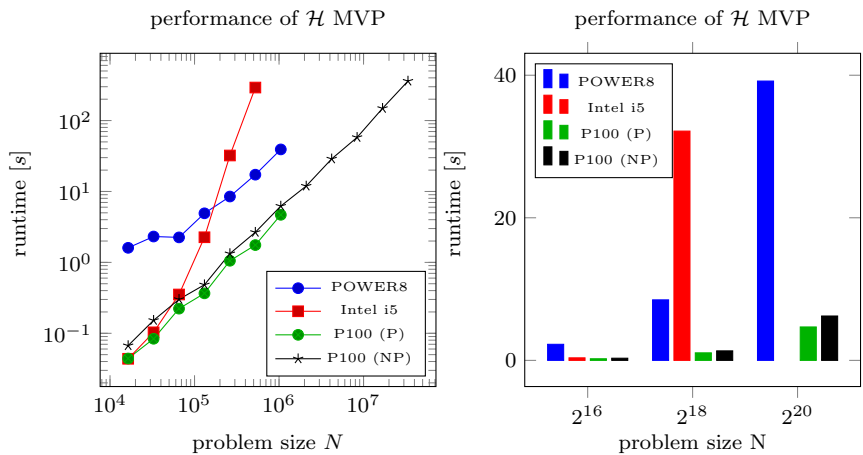

Our second comparative study targets the $\Hm$ matrix-vector product. It is done with the same parameters as before. The results are given in Fig.~\ref{sec:comparisonMVP}. While both GPU results and the POWER8-based result on CPU show the usual complexity behavior, we observe a significant increase in runtime for growing problem size on the Intel architecture. Right-now, the reason for this behavior is not clear. This is why we exclude this result from our discussion, here. Comparing the three remaining results, we observe a strong performance improvement of the GPU-based runs against the CPU-based results. Comparing again the results for $N=2^{19}$, we see a runtime of about 17 seconds on the CPU and a runtime of 2.7 seconds without ACA pre-computing and an improvement by about $60$ \% to 1.7 seconds with pre-computing on GPU. This is still a remarkable performance improvement by a factor of 10 on GPU. However, comparing a (fictive) 16-core parallel CPU implementation with the GPU results might result in a comparable performance between both architectures (depending on the scalability assumptions for the CPU). At this point, we still have to keep in mind that the CPU-based code assembles and stores all dense matrix sub-blocks of the approximated matrix, beforehand, while \texttt{hmglib} recomputes these on-the-fly due to memory limitations. Moreover, we still see some room for performance improvements of the GPU implementation.

Overall, we conclude that a perfectly fair comparison is hardly possible. CPU-based implementations rely much more on pre-computation and therefore might have a slight advantage for the $\Hm$ matrix-vector product in a 16-core CPU to GPU comparison, while being much slower in the setup phase. The new GPU-based implementation tries to balance the strong memory restrictions of GPUs with a general performance improvement. Based on the raw numbers of the single-threaded CPU to GPU comparison, the GPU code outperforms the CPU code by two orders of magnitude for the setup and by one order of magnitude for the $\Hm$ matrix-vector product.

\section{Summary}
\label{sec:summary}
This work considered the reformulation of algorithms in the construction and matrix-vector product of $\Hm$ matrices for many-core parallelism. As core techniques, to get fast parallel performance of $\Hm$ matrices on many-core hardware, we identified a parallel spatial data structure based on space filling curves, parallel tree traversal and batching of many small, non-equally sized compute tasks. On top of these basic building blocks, we designed algorithms for many-core parallel $\Hm$ matrices. These algorithms were transferred to a reference implementation on a GPU, which results in the GPU $\Hm$ matrix library \texttt{hmglib}. Our computation results section showed that the designed algorithms lead to a fast GPU implementation. Compared to the \textit{sequential} version of the \texttt{H2Lib} library, we achieve more than two orders of magnitude performance improvement on one Tesla P100 SXM2 GPU for the $\Hm$ matrix construction and roughly one order of magnitude performance in the $\Hm$ matrix-vector product. Note however that comparing the libraries and the underlying hardware is somewhat difficult. Nevertheless, we tried our best to keep this comparison fair.

In the future, our new algorithms shall be extended to the use in a dis\-tri\-bu\-ted-memory, thus e.g.~multi-GPU, context. This, however, involves to build an appropriate load balancing for the work distribution of ACA computations and dense matrix-vector products on an entire cluster of compute nodes equipped with many-core hardware. Moreover, the heterogeneous nature, i.e.~the existence of powerful CPUs and many-core devices, of current compute cluster should also be addressed, in order to get an even higher performance out of these systems.



\begin{acknowledgements}
This work is funded by the Swiss National Science Foundation (SNF) under project number 407540\_167186. 
Furthermore, code developments tasks in this research were done on resources of the Oak Ridge Leadership Computing Facility
at the Oak Ridge National Laboratory, which is supported by the Office of Science of the U.S. Department of Energy under Contract No. DE-AC05-00OR22725.
The IBM POWER8 system with the NVIDIA Tesla P100 SXM2 used in the benchmarks for this research was donated by the \textit{NVIDIA PSG Cluster}. All funding and support is gratefully acknowledged.

\end{acknowledgements}

\bibliographystyle{spmpsci}      
\bibliography{bibliography}   

\end{document}